\documentclass[sort&compress,10pt,twocolumn,final,3p]{elsarticle}

\usepackage[utf8]{inputenc}
\usepackage{hyperref}
\usepackage{graphicx}
\usepackage{subfigure}
\usepackage{array}
\usepackage{multirow}
\usepackage{amsmath,amssymb}
\usepackage{esvect}	
\usepackage{bm}	
\usepackage{xspace} 
\usepackage{textcomp} 

\newcommand{\etal}{\textit{et al}.\@\xspace}
\newcommand{\ie}{\textit{i.e.}\@\xspace}
\newcommand{\eg}{\textit{e.g.}\@\xspace}
\newcommand{\cf}{\textit{cf}.\@\xspace}

\newcommand{\abinitio}{\textit{ab initio}\@\xspace}

\newcommand{\ud}{\mathrm{d}}
\DeclareMathOperator{\bigO}{O}
\DeclareMathOperator{\dirac}{\delta}
\DeclareMathOperator{\Tr}{Tr}

\journal{Computational Materials Science}

\begin{document}

\begin{frontmatter}

\title{Elastic modeling of point-defects and their interaction}

\author[SRMP]{Emmanuel Clouet\corref{CA}}
\cortext[CA]{Corresponding author}
\ead{emmanuel.clouet@cea.fr}
\author[CiNAM]{Céline Varvenne}
\author[SRMP]{Thomas Jourdan}
\address[SRMP]{DEN-Service de Recherches de Métallurgie Physique, CEA,Paris-Saclay Univ., F-91191 Gif-sur-Yvette, France}
\address[CiNAM]{Centre Interdisciplinaire des Nanosciences de Marseille, UMR 7325 CNRS - Aix Marseille Univ., F-13008 Marseille, France}

\begin{abstract}
	Different descriptions used to model a point-defect 
	in an elastic continuum are reviewed.  
	The emphasis is put on the elastic dipole approximation, 
	which is shown to be equivalent to the infinitesimal Eshelby inclusion 
	and to the infinitesimal dislocation loop. 
	Knowing this elastic dipole, a second rank tensor fully characterizing 
	the point-defect, one can directly obtain the long-range elastic field induced by the point-defect 
	and its interaction with other elastic fields.
	The polarizability of the point-defect, resulting from the elastic dipole dependence 
	with the applied strain, is also introduced. 
	Parameterization of such an elastic model, either from experiments 
	or from atomic simulations, is discussed. 
	Different examples, like elastodiffusion and bias calculations,
	are finally considered to illustrate 
	the usefulness of such an elastic model 
	to describe the evolution of a point-defect in a external elastic field.
\end{abstract}


\end{frontmatter}

\section{Introduction}

Point-defects in crystalline solids, 
being either intrinsic like vacancies, self-interstitial atoms, 
and their small clusters,
or extrinsic like impurities and dopants, 
play a major role in materials properties and their kinetic evolution. 
Some properties of these point-defects, like their formation and migration energies, 
are mainly determined by the region in the immediate vicinity of the defect
where the crystal structure is strongly perturbed. 
An atomic description appears thus natural to model these properties,
and atomic simulations relying either on \abinitio calculations \cite{Freysoldt2014}
or empirical potentials have now become a routine tool to study point-defects 
structures and energies.
But point-defects also induce a long-range perturbation of the host lattice, 
leading to an elastic interaction with other structural defects, impurities
or an applied elastic field.
An atomic description thus appears unnecessary 
to capture the interaction arising from this long-range part,
and sometimes is also impossible 
because of the reduced size of the simulation cell in atomic approaches. 
Elasticity theory becomes then the natural framework.
It allows a quantitative description of the point-defect interaction
with other defects.

Following the seminal work of Eshelby \cite{Eshelby1956}, 
the simplest elastic model of a point-defect corresponds to 
a spherical inclusion forced into a spherical hole 
of slightly different size in an infinite elastic medium. 
This description accounts for the point-defect relaxation volume 
and its interaction with a pressure field (size interaction). 
It can be enriched by considering an ellipsoidal inclusion, 
thus leading to a interaction with also the deviatoric component of the stress field
(shape interaction),
and by assigning different elastic constants to the inclusion (inhomogeneity)
to describe the variations of the point-defect ``size'' and ``shape'' with the strain field 
where it is immersed. 
Other elastic descriptions of the point-defect are possible. 
In particular, it can be modeled by an equivalent distribution
of point-forces.  The long-range elastic field of the point-defect
and its interaction with other stress sources are then fully 
characterized by the first moment of this force distribution, 
a second-rank tensor called the elastic dipole. 
This description is rather natural when modeling point-defects 
and it can be used to extract elastic dipoles from atomic simulations.
These different descriptions are equivalent in the long-range limit, 
and allow for a quantitative modeling of the elastic field induced by the point-defect,
as long as the elastic anisotropy of the matrix is considered.

This article reviews these different elastic models which can be used to describe a point-defect 
and illustrates their usefulness with chosen examples.
After a short reminder of elasticity theory (Sec. \ref{sec:elasticity}),
we introduce the different descriptions of a point-defect within elasticity theory
(Sec. \ref{sec:point_defect}), favoring the elastic dipole description 
and showing its equivalence with the infinitesimal Eshelby inclusion
as well as with an infinitesimal dislocation loop. 
The next section (Sec. \ref{sec:para}) describes how the characteristics of the point-defect
needed to model it within elasticity theory can be obtained 
either from atomistic simulations or from experiments.
We finally give some applications in Sec. \ref{sec:examples}, 
where results of such an elastic model are compared to direct atomic simulations
to assess its validity.
The usefulness of this elastic description
is illustrated in this section for elastodiffusion and for the calculation of bias factors,
as well as for the modeling of isolated point-defects in atomistic simulations.

\section{Elasticity theory}
\label{sec:elasticity}

Before describing the modeling of a point-defect within elasticity theory, it is worth recalling the main aspects of the theory \cite{Landau1970}, 
in particular the underlying assumptions, some definitions and useful results.

\subsection{Displacement, distortion and strain}

Elasticity theory is based on a continuous description of solid bodies.   
It relates the forces, either internal or external, exerting on the solid to its deformation. 
To do so, one first defines the elastic displacement field.  
If $\vec{R}$ and $\vec{r}$ are the position of a point respectively in the unstrained and the strained body,
the displacement at this point is given by
\begin{equation*}
	\vec{u}(\vec{R}) = \vec{r} - \vec{R}.
\end{equation*}
One can then define the distortion tensor 
$\partial u_i \,/\, \partial R_j$
which expresses how an infinitesimal vector $\vv{\ud{R}}$ in the unstrained solid 
is transformed in $\vv{\ud{r}}$ in the strained body through the relation
\begin{equation*}
	\ud{r}_i = \left( \delta_{ij} + \frac{\partial u_i}{\partial R_j} \right) \ud{R}_j,	
\end{equation*}
where summation over repeated indices is implicit (Einstein convention)
and $\delta_{ij}$ is the Kronecker symbol.

Of central importance to the elasticity theory is the dimensionless strain tensor, 
defined by
\begin{align*}
	\varepsilon_{ij}(\vec{R}) 
	&= \frac{1}{2}\left[ \left( \delta_{in} + \frac{\partial u_n}{\partial R_i} \right)
		\left( \delta_{nj} + \frac{\partial u_n}{\partial R_j} \right) - \delta_{ij} \right] \\
	&= \frac{1}{2} \left( \frac{\partial u_i}{\partial R_j} + \frac{\partial u_j}{\partial R_i} 
		+ \frac{\partial u_n}{\partial R_i}\frac{\partial u_n}{\partial R_j} \right).
\end{align*}
This symmetric tensor expresses the change of size and shape of a body as a result of a force acting on it. 
The length $\ud{L}$ of the infinitesimal vector $\vv{\ud{R}}$ in the unstrained body
is  thus transformed into $\ud{l}$ in the strained body, through the relation
\begin{equation*}
	\ud{l}^2 = \ud{L}^2 + 2 \varepsilon_{ij} \ud{R}_i \ud{R}_j.
\end{equation*}

Assuming small deformation, a common assumption of linear elasticity, 
only the leading terms of the distortion are kept.  
The strain tensor then corresponds to the symmetric part of the distortion tensor, as
\begin{equation}
	\varepsilon_{ij}(\vec{R}) 
		= \frac{1}{2} \left( \frac{\partial u_i}{\partial R_j} + \frac{\partial u_j}{\partial R_i} \right).
	\label{eq:strain_def}
\end{equation}
The antisymmetric part of the distortion tensor corresponds to the infinitesimal rigid body rotation. 
It does not lead to any energetic contribution within linear elasticity in the absence of internal torque.

With this small deformation assumption, there is no distinction between
Lagrangian coordinates $\vec{R}$ and Eulerian coordinates $\vec{r}$
when describing elastic fields.
One can equally write, for instance, $\vec{u}(\vec{r})$ or $\vec{u}(\vec{R})$ for the displacement field,
which are equivalent to the leading order of the distortion.

\subsection{Stress}

The force $\vv{\delta F}$ acting on a volume element $\delta V$ of a strained body 
is composed of two contributions, 
the sum of external body forces $\vec{f}$
and the internal forces arising from atomic interactions. 
Because of the mutual cancellation of forces between particles inside the volume $\delta V$, 
only forces corresponding to the interaction with outside particles appear in this last contribution,
which is thus proportional to the surface elements $\vv{\ud S}$ defining the volume element $\delta V$. 
One obtains
\begin{equation*}
	\delta F_i = \int_{\delta V}{ f_i \ud{V}} + \oint_{\delta S}{\sigma_{ij}\ud S_j},
\end{equation*}
where $\sigma$ is the stress tensor defining internal forces.

Considering the mechanical equilibrium of the volume element $\delta V$, 
the absence of resultant force leads to the equation
\begin{equation}
	\frac{\partial \sigma_{ij}(\vec{r})}{\partial r_j} + f_i(\vec{r}) = 0,
	\label{eq:equil_stress}
\end{equation}
whereas the absence of torque ensures the symmetry of the stress tensor.

At the boundary of the strained body, internal forces are balanced by applied forces. 
If $\vec{T}^{\rm a} \ud{S}$ is the force applied on the infinitesimal surface element $\ud{S}$, 
this leads to the boundary condition
\begin{equation}
	\sigma_{ij} n_j = T^{\rm a}_i, 
	\label{eq:equil_stress_boundary}
\end{equation}
where $\vec{n}$ is the outward-pointing normal to the surface element $\ud{S}$.

The work $\delta w$, defined per volume unit, of these internal forces is given by
\begin{equation*}
	\delta w = -\sigma_{ij} \delta{\varepsilon_{ij}},
\end{equation*}
where $\delta \varepsilon_{ij}$ is the strain change during the deformation increase, 
and the sign convention is $\delta w > 0$ when the energy flux goes
outwards the elastic body.
This leads to the following thermodynamic definition of the stress tensor
\begin{equation*}
	\sigma_{ij} = \left( \frac{\partial e}{\partial \varepsilon_{ij}} \right)_s
	            = \left( \frac{\partial f}{\partial \varepsilon_{ij}} \right)_T,
\end{equation*}
where $e$, $s$, and $f=e-Ts$ are the internal energy, entropy, and free energy of the elastic body
defined per volume unit.

\subsection{Hooke's law}
\label{sec:elast_Hooke}

To go further, one needs a constitutive equation for the energy or the free energy.
Taking as a reference the undeformed state corresponding to the elastic body
at equilibrium without any external force, either body or applied stress, 
the energy is at a minimum for $\varepsilon=0$ and then
\begin{equation*}
	\sigma_{ij}(\varepsilon=0) 
		= \left.  \frac{\partial e}{\partial \varepsilon_{ij}} \right|_{\varepsilon = 0} 
		= 0.
\end{equation*}
The leading order terms of the series expansion of the energy are then
\begin{equation*}
	e(T,\varepsilon) = e^0(T) + \frac{1}{2} C_{ijkl}\varepsilon_{ij}\varepsilon_{kl},
\end{equation*}
where $e^0(T) = e(T,\varepsilon=0)$ is the energy of the unstrained body at temperature $T$.
The elastic constants $C_{ijkl}$ entering this expression are thus defined by
\begin{equation*}
	C_{ijkl} = \frac{\partial^2 e}{\partial \varepsilon_{ij} \partial \varepsilon_{kl}}.
\end{equation*}
This is a fourth-rank tensor which obeys 
minor symmetry $C_{ijkl}=C_{jikl}=C_{ijlk}$ because of the strain tensor symmetry 
and also major symmetry $C_{ijkl}=C_{klij}$ because of allowed permutation of partial derivatives.
This leads to at most 21 independent coefficients, which can be further reduced 
by considering the symmetries of the solid body \cite{Nye1957}.

This series expansion of the energy leads to a linear relation, the Hooke's law,
between the stress and the strain
\begin{equation}
	\sigma_{ij} = C_{ijkl} \varepsilon_{kl},
	\label{eq:stress_Hooke}
\end{equation}
which was summarized in 1678 by Robert Hooke 
as \emph{Ut tensio, sic vis}.\footnote{As the extension, so the force.}

\subsection{Elastic equilibrium, superposition principle}

Combining Hooke's law \eqref{eq:stress_Hooke} with 
the small deformation definition \eqref{eq:strain_def} of the strain tensor
and the equilibrium condition \eqref{eq:equil_stress},
one obtains the equation obeyed by the displacement at equilibrium
\begin{equation}
	C_{ijkl} \frac{\partial^2 u_k(\vec{r})}{\partial r_j \partial r_l} + f_i(\vec{r}) = 0.
	\label{eq:equil_displacement}
\end{equation}
The elastic equilibrium is given by the solution 
which verifies the boundary conditions, 
$\sigma_{ij}n_ j = T_i^{\rm a}$ for imposed applied forces
and $u_i=u_i^{\rm a}$ for imposed applied displacements.

As elastic equilibrium is defined by the solution of a linear partial differential equation (Eq. \ref{eq:equil_displacement}),
the superposition principle holds.  
If two elastic fields, characterized by their displacement $\vec{u}^1(\vec{r})$ and  $\vec{u}^2(\vec{r})$,
correspond to equilibrium for the respective body forces $\vec{f}^1$ and $\vec{f}^2$ 
and the respective boundary conditions $(\vec{u}^{\rm a1}, \vec{T}^{\rm a1})$ and  $(\vec{u}^{\rm a2}, \vec{T}^{\rm a2})$,
then the elastic equilibrium for the body forces $\vec{f}^1 + \vec{f}^2$
and the boundary conditions $(\vec{u}^{\rm a1} + \vec{u}^{\rm a2}, \vec{T}^{\rm a1} + \vec{T}^{\rm a2})$
is given by the sum of these two elastic fields.
The total elastic energy is composed of the contributions of each elastic field taken separately 
and an interaction energy given by
\begin{equation}
	\begin{split}
	E^{\rm int} 
	=& \int_{V}{ \sigma_{ij}^1(\vec{r}) \, \varepsilon_{ij}^2(\vec{r}) \, \ud{V} } \\
	=& \int_{V}{ \sigma_{ij}^2(\vec{r}) \, \varepsilon_{ij}^1(\vec{r}) \, \ud{V} }.
	\end{split}
	\label{eq:Eqinter}
\end{equation}
This equation can be used to define interaction energy between two defects.

The superposition principle allows making use of Green's function.  
The elastic Green's function $G_{kn}(\vec{r})$ is the solution of the equilibrium equation 
for a unit point-force
\begin{equation}
	C_{ijkl} \frac{\partial^2 G_{kn}(\vec{r})}{\partial r_j \partial r_l} + \delta_{in} \dirac(\vec{r}) = 0,
	\label{eq:equil_Green}
\end{equation}
where $\dirac(\vec{r})$ is the Dirac delta function, 
\ie $\delta(\vec{r})=0$ if $\vec{r}\neq\vec{0}$
and $\delta(\vec{0})=\infty$.
$G_{kn}(\vec{r})$ therefore corresponds to the displacement along the $r_{k}$ axis 
for a unit point-force applied along the $r_{n}$ axis at the origin.
The solution of elastic equilibrium for the force distribution $\vec{f}(\vec{r})$
is then given by
\begin{align*}
	u_k(\vec{r}) 		&= \int_V{ G_{kn}( \vec{r} - \vec{r}^{\,\prime} ) f_n( \vec{r}^{\,\prime} )  \ud{V^{\,\prime}} }, \\
	\sigma_{ij}(\vec{r}) 	&= C_{ijkl} \int_V{ G_{kn,l}( \vec{r} - \vec{r}^{\,\prime} ) f_n( \vec{r}^{\,\prime} )  \ud{V^{\,\prime}} }, 
\end{align*}
where we have introduced the notation $G_{kn,l} = \partial G_{kn} \,/\, \partial r_l$
for partial derivatives.

An analytical expression of the Green's function exists for isotropic elasticity.
Considering the elastic constants $C_{ijkl} = \lambda \delta_{ij}\delta_{kl} + \mu( \delta_{ik}\delta_{jl}+\delta_{il}\delta_{jk})$,
where $\lambda$ and $\mu$ are the Lamé coefficients, 
the Green's function is given by
\begin{equation*}
	G_{kn}(\vec{r}) = \frac{1}{ 8 \pi \mu} \left[ \frac{\lambda+3\mu}{\lambda+2\mu} \delta_{kn} + \frac{\lambda+\mu}{\lambda+2\mu} \eta_k \eta_n \right] \frac{1}{r},
\end{equation*}
with $r = \| \vec{r} \|$ and $\vec{\eta}=\vec{r}/r$.
No analytical expression exists in the more general case of elastic anisotropy,
but the Green's function, and its successive derivatives, can be calculated efficiently 
from the elastic constants using the numerical scheme of Barnett \cite{Barnett1972a,Bacon1980}.
Whatever the anisotropy, the Green's function and its derivatives will show the same variation 
with the distance $r$,\footnote{The scaling with the distance $r$ is a consequence of Eq. \eqref{eq:equil_Green},
given that the $\delta(\vec{r})$ function is homogeneous of degree $-3$.}
leading to the general expressions
\begin{equation*}
	G_{kn}(\vec{r}) = g_{kn}(\vec{\eta}) \frac{1}{r}
	\textrm{\ , }
	G_{kn,l}(\vec{r}) = h_{knl}(\vec{\eta}) \frac{1}{r^2}
	\textrm{\ , \dots }
\end{equation*}
where the anisotropy enters only in the angular dependence $g_{kn}(\vec{\eta})$,  $h_{knl}(\vec{\eta})$, \dots

\section{Elastic model of a point-defect}
\label{sec:point_defect}

Different models can be used to describe a point-defect within elasticity theory. 
One such model is the elastic dipole.  We first describe this model 
and then demonstrate the analogy with a description of the point-defect
as an infinitesimal Eshelby inclusion or an infinitesimal dislocation loop.
We finally introduce the polarizability of the point-defect.

\subsection{Elastic dipole}
\label{sec:dipole_model}

A point-defect can be described in a continuous solid body as an equilibrated distribution 
of point-forces \cite{Siems1968,Leibfried1978,Bacon1980,Teodosiu1982}.  
Considering a point-defect located at the origin modeled by such a force distribution  
$\vec{f}(\vec{r}) = \sum_{q=1}^N{ \vec{F}^q \dirac{(\vec{r}-\vec{a}^q})}$,  
\ie consisting of $N$ forces $\vec{F}^q$ each acting at position $\vec{a}^q$, 
the elastic displacement field of the point-defect is, according to linear elasticity theory,
given by
\begin{equation*}
	u_i(\vec{r}) = \sum_{q=1}^N{ G_{ij}( \vec{r} - \vec{a}^q ) F^q_j },
\end{equation*}
where we have used the elastic Green's function.
Far from the point-defect, we have $\| \vec{r} \| \gg \| \vec{a}^q \|$ 
and we can make a series expansion of the Green's function:
\begin{multline*}
	u_i(\vec{r}) = 
		G_{ij}( \vec{r} ) \sum_{q=1}^N{ F^q_j }
		\ - \  G_{ij,k}( \vec{r} ) \sum_{q=1}^N{ F^q_j a^q_k }
		\\
		\ + \  \bigO{\left( \| \vec{a}^q \|^2 \right)} .
\end{multline*}
As the force distribution is equilibrated, its resultant 
$\sum_q{\vec{F}^q}$ is null. 
The displacement is thus given, to the leading order, by
\begin{equation}
	u_i(\vec{r}) = - G_{ij,k}( \vec{r} ) P_{jk},
	\label{eq:dipole_displacement}
\end{equation}
and the corresponding stress field by 
\begin{equation}
	\sigma_{ij}(\vec{r}) = - C_{ijkl} G_{km,nl}( \vec{r} ) P_{mn},
	\label{eq:dipole_stress}
\end{equation}
where the elastic dipole is defined as the first moment of the point-force distribution,
\begin{equation}
	P_{jk} = \sum_{q=1}^N{ F^q_j a^q_k }.
	\label{eq:dipole}
\end{equation}
This dipole is a second rank tensor which fully characterizes 
the point-defect within elasticity theory \cite{Siems1968,Leibfried1978,Bacon1980,Teodosiu1982}.
It is symmetric because the torque $\sum_q{ \vec{F}^q \times \vec{a}^q }$ 
must be null for the force distribution to be equilibrated.

Equations \eqref{eq:dipole_displacement} and \eqref{eq:dipole_stress} show
that the elastic displacement and the stress created by a point-defect 
are long-ranged, respectively decaying as $1/r^2$ and $1/r^3$ with the distance $r$
to the point-defect.

The elastic dipole is directly linked to the point-defect relaxation volume.
Considering a finite volume $V$ of external surface $S$ enclosing the point-defect, 
this relaxation volume is defined as
\begin{equation*}
	\Delta V = \oint_S { u_i(\vec{r}) \, \ud{S_i} },
\end{equation*}
where $\vec{u}(\vec{r})$ is the superposition of the displacement 
created by the point-defect (Eq. \ref{eq:dipole_displacement})
and the elastic displacement due to image forces ensuring null tractions on the external surface $S$. 
Use of the Gauss theorem, of the equilibrium condition \eqref{eq:equil_displacement} 
and of the elastic dipole definition \eqref{eq:dipole} leads to the result \cite{Leibfried1978}
\begin{equation}
	\Delta V =   S_{iikl} P_{kl},
	\label{eq:dipole_relax_vol}
\end{equation}
where the elastic compliances $S_{ijkl}$ are the inverse of the elastic constants,
\ie $S_{ijkl}C_{klmn} = \frac{1}{2}( \delta_{im}\delta_{jn} + \delta_{in}\delta_{jm} )$.
For a crystal with cubic symmetry, this equation can be further simplified \cite{Leibfried1978}
to show that the relaxation volume is equal to the trace of the elastic dipole divided 
by three times the bulk modulus.
More generally, as it will become clear with the comparison to the Eshelby's inclusion, 
this elastic dipole is the source term defining 
the relaxation volume of the point-defect. 
Its trace gives rise to the size interaction, 
whereas its deviator, \ie the presence of off-diagonal terms 
and differences in the diagonal components, leads to the shape interaction.

Of particular importance is the interaction energy of the point-defect with an external elastic field
$\vec{u}^{\rm ext}(\vec{r})$. 
Considering the point-forces distribution representative of the point-defect,
this interaction energy can be simply written as \cite{Bacon1980}
\begin{equation*}
	E^{\rm int} = -\sum_{q=1}^N{ F_i^q \, u_i^{\rm ext}(\vec{a}^q)}. 
\end{equation*}
If we now assume that the external field is slowly varying close to the point-defect, 
one can make a series expansion of the corresponding displacement $\vec{u}^{\rm ext}(\vec{r})$.
The interaction energy is then, to first order,
\begin{equation*}
	E^{\rm int} = - u_i^{\rm ext}(\vec{0}) \sum_{q=1}^N{ F_i^q } 
	- u_{i,j}^{\rm ext}(\vec{0}) \sum_{q=1}^N{ F_i^q \, a_j^q}. 
\end{equation*}
Finally, using the equilibrium properties of the point-forces distribution,
one obtains
\begin{equation}
	E^{\rm int} = - P_{ij} \, \varepsilon^{\rm ext}_{ij}(\vec{0}),
	\label{eq:dipole_Einter}
\end{equation}
thus showing that the interaction energy is simply the product of the elastic dipole
with the value at the point-defect location of the external strain field.
Higher order contributions to the interaction energy involve successive gradients of the external strain field
coupled with higher moments of the multipole expansion of the force distribution, 
and can be generally safely ignored.
This simple expression of the interaction energy is the workhorse of the modeling of point-defects 
within linear elasticity in a multiscale approach.

Instead of working with the elastic dipole tensor, 
one sometimes rather uses the so-called $\lambda$-tensor \cite{Nowick1972}
which expresses the strain variation of a matrix volume 
with the point-defect volume concentration $c$,
\begin{equation}
	\lambda_{ij} = \frac{1}{\Omega_{\rm at}} \, \frac{\partial \bar{\varepsilon}_{ij}}{\partial c},
	\label{eq:lambda_PD}
\end{equation}
where $\bar{\varepsilon}$ is the homogeneous strain 
induced by the point-defects in a stress-free state
and $\Omega_{\rm at}$ is the atomic volume of the reference solid.
As it will become clear when discussing parameterization of the elastic dipole 
from experiments (\S \ref{sec:para_exp}), these two quantities are simply linked
by the relation
\begin{equation}
	P_{ij} = \Omega_{\rm at} \, C_{ijkl} \, \lambda_{kl}.
	\label{eq:dipole_lamba}
\end{equation}
Using this $\lambda$-tensor to characterize the point-defect, 
Eq. \eqref{eq:dipole_Einter} describing its elastic interaction with an external elastic field 
becomes
\begin{equation*}
	E^{\rm int} = - \Omega_{\rm at} \, \lambda_{ij} \, \sigma_{ij}^{\rm ext}(\vec{0}),
\end{equation*}
where $\sigma^{\rm ext}_{ij}(\vec{0})$ is the value of the external stress field at the point-defect position.

\subsection{Analogy with Eshelby's inclusion}

The Eshelby's inclusion \cite{Eshelby1957,Eshelby1959a} is another widespread model
which can be used to describe a point-defect in an elastic continuum.
As it will be shown below, it is equivalent to the dipole description in the limit of an infinitesimal inclusion.

In this model, the point-defect is described as an inclusion of volume $\Omega_{\rm I}$ and of surface $S_{\rm I}$,
having the same elastic constants as the matrix. This inclusion undergoes a change of shape 
described by the eigenstrain $\varepsilon_{ij}^*(\vec{r})$, corresponding to the strain that would adopt the inclusion 
if it was free to relax and was not constrained by the surrounding matrix.
Eshelby proposed a general approach \cite{Eshelby1957} to solve the corresponding equilibrium problem
and determine the elastic fields in the inclusion and the surrounding matrix.
This solution is obtained by considering the three following steps:
\begin{enumerate}
	\item Take the inclusion out of the matrix and let it adopt its eigenstrain $\varepsilon_{ij}^*(\vec{r})$.
		At this stage, the stress is null everywhere.
	\item Strain back the inclusion so it will fit the hole in the matrix. 
		The elastic strain exactly compensates for the eigenstrain, 
		so the stress in the inclusion is $-C_{ijkl}\varepsilon^*_{kl}(\vec{r})$.
		This operation is performed by applying to the external surface of the inclusion 
		the traction forces corresponding to this stress
		\begin{equation*}
			\ud{T}_i(\vec{r}) = -C_{ijkl} \, \varepsilon^*_{kl}(\vec{r}) \, \ud{S}_j,
		\end{equation*}
		where $\vv{\ud{S}}$ is an element of the inclusion external surface at the point $\vec{r}$.
	\item After the inclusion has been welded back into its hole, the traction forces are relaxed. 
		Using Green's function, the corresponding displacement in the matrix is then
		\begin{equation*}
			\begin{split}
				u_n(\vec{r}) 
				&=  \oint_{S_{\rm I}} { G_{ni}(\vec{r}-\vec{r}^{\,\prime}) \, \ud{T}_i(\vec{r}^{\,\prime})}, \\
				&= -\oint_{S_{\rm I}} { G_{ni}(\vec{r}-\vec{r}^{\,\prime}) \, C_{ijkl} \, \varepsilon^*_{kl}(\vec{r}^{\,\prime}) \, \ud{S}_j^{\,\prime}}.
			\end{split}
		\end{equation*}
\end{enumerate}
Applying Gauss theorem and the equilibrium condition satisfied by the eigenstrain 
$\varepsilon_{ij}^*(\vec{r})$,
one obtains the following expression for the elastic displacement in the matrix
\begin{equation}
	u_n(\vec{r}) =  -\int_{\Omega_{\rm I}} { G_{ni,j}(\vec{r}-\vec{r}^{\,\prime}) \, C_{ijkl} \, \varepsilon^*_{kl}(\vec{r}^{\,\prime}) \, \ud{V^{\,\prime}} },
	\label{eq:inclusion_displacement}
\end{equation}
and for the corresponding stress field
\begin{multline}
	\sigma_{pq}(\vec{r}) =  -\int_{\Omega_{\rm I}}{ C_{pqmn} \, G_{ni,jm}(\vec{r}-\vec{r}^{\,\prime}) } \\
		{ C_{ijkl} \, \varepsilon^*_{kl}(\vec{r}^{\,\prime}) \, \ud{V^{\,\prime}} }.
	\label{eq:inclusion_stress}
\end{multline}
Inside the inclusion, one needs to add the stress $-C_{ijkl}\varepsilon^*_{kl}(\vec{r})$
corresponding to the strain applied in step 2.

Far from the inclusion, we have $\| \vec{r} \| \gg \| \vec{r}^{\,\prime} \|$.
We can therefore neglect the variations of the Green's function derivatives 
inside Eqs. \ref{eq:inclusion_displacement} and \ref{eq:inclusion_stress}. 
This corresponds to the infinitesimal inclusion assumption. 
For such an infinitesimal inclusion located at the origin, one therefore obtains the following elastic fields
\begin{align}
	u_n(\vec{r}) &=  - G_{ni,j}(\vec{r}) \, C_{ijkl} \, \Omega_{\rm I} \, \bar{\varepsilon}^*_{kl},
	\label{eq:small_inclusion_displacement}
	\\
	\sigma_{pq}(\vec{r}) &=  - C_{pqmn} \, G_{ni,jm}(\vec{r}) \,
		C_{ijkl} \, \Omega_{\rm I} \, \bar{\varepsilon}^*_{kl},
	\label{eq:small_inclusion_stress}
\end{align}
where we have defined the volume average of the inclusion eigenstrain,
$\bar{\varepsilon}_{ij}^* = \frac{1}{\Omega_{\rm I}}\int_{\Omega_{\rm I}}{ \varepsilon_{ij}(\vec{r}) \, \ud{V}}$.
Comparing these expressions with the ones describing the elastic field of an elastic dipole
(Eqs. \ref{eq:dipole_displacement} and \ref{eq:dipole_stress}), we see that they are the same
for any $\vec{r}$ value
provided the dipole tensor and the inclusion eigenstrain check the relation
\begin{equation}
	P_{ij} = \Omega_{\rm I} \, C_{ijkl} \, \bar{\varepsilon}_{kl}^*.
	\label{eq:small_inclusion_dipole}
\end{equation}
The descriptions of a point-defect as an elastic dipole,
\ie as a distribution of point-forces keeping only the first moment of the distribution,
or as an infinitesimal Eshelby  inclusion, 
\ie in the limit of an inclusion volume $\Omega_{\rm I}\to 0$ 
keeping the product $\Omega_{\rm I}\,\bar{\varepsilon}_{ij}^*$ constant,
are therefore equivalent.
The point-defect can be thus characterized either by its elastic dipole tensor $P_{ij}$
or by its eigenstrain tensor $Q_{ij}=\Omega_{\rm I}\,\bar{\varepsilon}_{ij}^*$ \cite{Lazar2017}.

Of course, the same equivalence is obtained when considering the interaction energy with an external stress field.
For a general inclusion, Eshelby showed that this interaction energy is simply given by
\begin{equation}
	E^{\rm int} = -\int_{\Omega_{\rm I}}{\varepsilon^*_{ij}(\vec{r}) \, \sigma_{ij}^{\rm ext}(\vec{r}) \, \ud{V} },
	\label{eq:inclusion_Einter}
\end{equation}
where the integral only runs on the inclusion volume.
In the limiting case of an infinitesimal inclusion, one can neglect the variations of the external stress field
inside the inclusion.  One thus obtains the following interaction energy,
\begin{equation}
	E^{\rm int} = - \Omega_{\rm I} \, \bar{\varepsilon}^*_{ij} \, \sigma_{ij}^{\rm ext}(\vec{0}) ,
	\label{eq:small_inclusion_Einter}
\end{equation}
which is equivalent to the expression \eqref{eq:dipole_Einter} for an elastic dipole  
when the equivalence relation \eqref{eq:small_inclusion_dipole} is verified.

\subsection{Analogy with dislocation loops}

A point-defect can also be considered as an infinitesimal dislocation loop.
This appears natural as dislocation loops are known to be elastically equivalent to platelet Eshelby's inclusions 
\cite{Nabarro1967,Mura1987}.

The elastic displacement and stress fields of a dislocation loop of Burgers vector $\vec{b}$
are respectively given by the Burger's and the Mura's formulae \cite{Hirth1982}
\begin{align}
	\begin{split}
		u_i(\vec{r}) ={ }& 
			C_{jklm} \,  b_m \\
			& \quad 
			\int_{A}{ G_{ij,k}(\vec{r} - \vec{r}^{\,\prime}) \, n_l(\vec{r}^{\,\prime}) \, \ud{A^{\,\prime}} },
		\label{eq:dislo_loop_displacement}
	\end{split}
	\\
	\begin{split}
		\sigma_{ij}(\vec{r}) ={ }&
			C_{ijkl} \, \epsilon_{lnh} C_{pqmn} b_m \\
			& \quad
			\oint_{L}{ G_{kp,q}(\vec{r} - \vec{r}^{\,\prime})  \, \zeta_h(\vec{r}^{\,\prime}) \, \ud{l^{\,\prime}} }.
	\end{split}
	\label{eq:dislo_loop_stress}
\end{align}
The displacement is defined by a surface integral on the surface $A$ enclosed by the dislocation loop,
with $\vec{n}(\vec{r}^{\,\prime})$ the local normal to the surface element $\ud{A^{\,\prime}}$ in $\vec{r}^{\,\prime}$,
and the stress by a line integral along the loop of total line length $L$.  
$\vec{\zeta}$ is the unit vector along the loop,
and $\epsilon_{lnh}$ is the permutation tensor.

Like for the Eshelby's inclusion, far from the loop ($\| \vec{r} \| \gg \| \vec{r}^{\,\prime} \|$), 
we can use a series expansion of the Green's function derivatives and keep only the leading term. 
Considering a loop located at the origin, we thus obtain
\begin{align}
	u_i(\vec{r}) =& C_{jklm} \,  b_m \, A_l \, G_{ij,k}(\vec{r}) ,
	\label{eq:small_dislo_loop_displacement}
	\\
	\sigma_{pq}(\vec{r}) =& C_{pqin} \, C_{jklm} \, b_m \, A_l \, G_{ij,kn}(\vec{r}) ,
	\label{eq:small_dislo_loop_stress}
\end{align}
where $\vec{A}$ is the surface vector defining the area of the loop.
These expressions are equal to the ones obtained for an elastic dipole \eqref{eq:dipole_displacement} and
\eqref{eq:dipole_stress}, with the equivalent dipole tensor of the dislocation loop given by
\begin{equation}
	P_{jk} = - C_{jklm} \,  b_m \, A_l.
	\label{eq:small_dislo_loop_dipole}
\end{equation}

Looking at the interaction with an external stress field, 
the interaction energy with the dislocation loop is given by
\begin{equation}
	E^{\rm int} = \int_{A}{ \sigma_{ij}^{\rm ext}(\vec{r}) \, b_i \, n_j \, \ud{A} }.
	\label{eq:dislo_loop_Einter}
\end{equation}
For an infinitesimal loop, it simply becomes
\begin{equation}
	E^{\rm int} = \sigma_{ij}^{\rm ext}(\vec{0}) \, b_i \, A_j,
	\label{eq:small_dislo_loop_Einter}
\end{equation}
which is equivalent to the expression \eqref{eq:dipole_Einter} obtained for an elastic dipole
when the equivalent dipole tensor of the dislocation loop is given by Eq. \eqref{eq:small_dislo_loop_dipole}.

\subsection{Polarizability}

The equivalent point-forces distribution of a point-defect can be altered by an applied elastic field \cite{Kroner1964}. 
This applied elastic field thus leads to an induced elastic dipole 
and the total elastic dipole of the point-defect now depends on the applied strain $\varepsilon^{\rm ext}$:
\begin{equation}
	P_{ij}(\varepsilon^{\rm ext}) = P_{ij}^{0} + \alpha_{ijkl} \varepsilon^{\rm ext}_{kl},
	\label{eq:polarizability}
\end{equation}
where $P_{ij}^{0}$ is the permanent elastic dipole in absence of applied strain 
and $\alpha_{ijkl}$ is the point-defect diaelastic polarizability \cite{Schober1984,Puls1986,Granato1994}. 
Considering the analogy with the Eshelby's inclusion,
this polarizability corresponds to an infinitesimal inhomogeneous inclusion, 
\ie an inclusion with different elastic constants than the surrounding matrix.
It describes the fact that the matrix close to the point-defect has a different elastic response 
to an applied strain because of the perturbations of the atomic bonding caused by the point-defect.
For the analogy with an infinitesimal dislocation loop, 
the polarizability corresponds to the fact that the loop can change its shape by glide on its prismatic cylinder
(or in its habit plane for a pure glide loop) under the action of the applied elastic field.

Following Schober \cite{Schober1984}, the interaction of a point-defect located at the origin with an applied strain 
is now given by
\begin{equation}
	E^{\rm int} = -P^0_{ij} \, \varepsilon^{\rm ext}_{ij}(\vec{0}) 
		- \frac{1}{2} \, \alpha_{ijkl} \, \varepsilon^{\rm ext}_{ij}(\vec{0}) \, \varepsilon^{\rm ext}_{kl}(\vec{0}).
	\label{eq:dipole_polar_Einter}
\end{equation}
This expression of the interaction energy, which includes the defect polarizability, has important consequences 
for the modeling of point-defects as it shows that some coupling is possible between two different applied elastic fields.
Considering the point-defect interaction with the two strain fields $\varepsilon^{(1)}$ and $\varepsilon^{(2)}$
originating from two different sources, the interaction energy is now given by
\begin{align*}
	\begin{split}
		E^{\rm int} ={ }& -P^0_{ij} \left( \varepsilon^{(1)}_{ij} +  \varepsilon^{(2)}_{ij} \right) \\
			&\qquad - \frac{1}{2} \, \alpha_{ijkl} 
				\left( \varepsilon^{(1)}_{ij} +  \varepsilon^{(2)}_{ij} \right)
				\left( \varepsilon^{(1)}_{kl} +  \varepsilon^{(2)}_{kl} \right),
	\end{split}
	\\
	\begin{split}
		 ={ }& -P^0_{ij}  \varepsilon^{(1)}_{ij} - \frac{1}{2} \, \alpha_{ijkl} \, \varepsilon^{(1)}_{ij} \, \varepsilon^{(1)}_{kl} \\
		 &\qquad -P^0_{ij}  \varepsilon^{(2)}_{ij} - \frac{1}{2} \, \alpha_{ijkl} \, \varepsilon^{(2)}_{ij} \, \varepsilon^{(2)}_{kl} \\
		 &\qquad \qquad - \alpha_{ijkl} \, \varepsilon^{(1)}_{ij} \, \varepsilon^{(2)}_{kl}.
	\end{split}
\end{align*}
The last line therefore shows that, without the polarizability, 
the interaction energy of the point-defect with the two strain fields 
will be simply the superposition of the two interaction energies with each strain fields considered separately.
A coupling is introduced only through the polarizability.
Such a coupling is for instance at the origin of one of the mechanisms proposed to explain creep under irradiation. 
Indeed, because of the polarizability, the interaction of point-defects, either vacancies or self-interstitial atoms,
with dislocations under an applied stress
depends on the dislocation orientation with respect to the applied stress. 
This stronger interaction with some dislocation families leads to a larger drift term in the diffusion equation 
of the point-defect and thus to a greater absorption of the point-defect by these dislocations,
a mechanism known as Stress Induced Preferential Absorption (or SIPA) \cite{Heald1974,Heald1975b,Bullough1975a,Bullough1975b}.
This polarizability is also the cause, in alloy solid solutions,
of the variation of the matrix elastic constants with their solute content.

This diaelastic polarizability caused by the perturbation of the elastic response of the surrounding matrix manifests itself 
at the lowest temperature, even 0\,K, and whatever the characteristic time of the applied strain.  
At finite temperature there may be another source of polarizability.  
If the point-defect can adopt different configurations, 
for instance different variants corresponding to different orientations of the point-defect 
like for a carbon interstitial atom in a body-centered cubic Fe matrix,
then the occupancy distribution of these configurations will be modified under an applied stress or strain. 
This possible redistribution of the point-defect 
gives rise to anelasticity \cite{Nowick1972}, 
the most famous case being the Snoek relaxation in iron alloys containing interstitial solute atoms
like C and N \cite{Snoek1941}.
When thermally activated transitions between the different configurations of the point-defect are fast enough 
compared to the characteristic time of the applied stress, the distribution of the different configurations 
corresponds to thermal equilibrium.
Assuming that all configurations have the same energy in a stress-free state 
and denoting by $P^{\mu}_{ij}$ the elastic dipole of the configuration $\mu$, 
the average dipole of the point-defect is then given by
\begin{equation*}
	\langle P_{ij} \rangle = \frac{ \sum_{\mu}{ \exp{\left( P_{kl}^{\mu}\varepsilon_{kl}^{\rm ext}\,/\,kT \right)} P_{ij}^{\mu} } }
	{ \sum_{\mu}{ \exp{\left( P_{kl}^{\mu}\varepsilon_{kl}^{\rm ext}\,/\,kT \right)} } }.
\end{equation*}
As a consequence, the average elastic dipole of the point-defect distribution is now depending 
on the applied stress and on the temperature, an effect known as paraelasticity \cite{Kroner1964}.
At temperatures high enough to allow for transition between the different configurations, 
the interaction energy of the configurations with the applied strain 
is usually small compared to $kT$.
One can make a series expansion of the exponentials to obtain
\begin{equation*}
	\begin{split}
	\langle P_{ij} \rangle &= \frac{1}{n_{\mathrm{v}}}\sum_{\mu=1}^{n_{\mathrm{v}}}{ P_{ij}^{\mu} } \\
	&- \left(
	\frac{1}{ {n_{\rm v}}^2}\sum_{\mu, \nu=1}^{n_{\mathrm{v}}}{ P_{ij}^{\mu} P_{kl}^{\nu} } 
	- \frac{1}{n_{\mathrm{v}}}\sum_{\mu=1}^{n_{\mathrm{v}}}{ P_{ij}^{\mu} P_{kl}^{\mu} }
	\right)
	\frac{ \varepsilon_{kl}^{\rm ext} }{kT},
	\end{split}
\end{equation*}
where $n_{\mathrm{v}}$ is the number of configurations. 
This leads to the same linear variation of the elastic dipole with the applied strain 
as for the diaelastic polarizability (Eq. \ref{eq:polarizability}), 
except that the paraelastic polarizability is depending on the temperature.

\section{Parameterization of elastic dipoles}
\label{sec:para}

To properly model a point-defect with continuum elasticity theory,
one only needs to know its elastic dipole.
It is then possible to describe the elastic displacement (Eq. \ref{eq:dipole_displacement})
or the stress field (Eq. \ref{eq:dipole_stress}) induced by the point-defect, 
and also to calculate its interaction with an external elastic field (Eq. \ref{eq:dipole_Einter}).
This elastic dipole can be determined either using atomistic simulations
or from experiments.

\subsection{From atomistic simulations}
\label{sec:para_atom}

Different strategies can be considered for the identification of elastic dipoles in atomistic simulations. 
This elastic dipole can be directly deduced from the stress existing in the simulation box, 
or from a fit of the atomic displacements, 
or finally from a summation of the Kanzaki forces. 
We examine here these three techniques and discuss their merits and drawbacks.

\subsubsection*{Definition from the stress}

Let us consider a simulation box of volume $V$, the equilibrium volume of the pristine bulk material.
We introduce one point-defect in the simulation box and assume periodic boundary conditions 
to preclude any difficulty associated with surfaces.
Elasticity theory can be used to predict the variation of the energy 
of the simulation box submitted to a homogeneous strain $\varepsilon$. 
Using the interaction energy of a point-defect with an external strain given in Eq.~\eqref{eq:dipole_Einter},
one obtains
\begin{equation}
E(\varepsilon)  =  E_0 + E^{\rm PD} + \frac{V}{2}C_{ijkl}\varepsilon_{ij}\varepsilon_{kl} - P_{ij}\varepsilon_{ij},
\label{eq:energy_box_PD}
\end{equation}
with $E_0$ the bulk reference energy and $E^{\rm PD}$ the point-defect energy,
which can contain a contribution from the interactions of the point-defect with its periodic images
(see section~\ref{sec:elast_corr}). 
The average residual stress on the simulation box is obtained by simple derivation as\footnote{See also
Refs. \cite{Puchala2008} and \cite{Pasianot2016b} for other proofs.}
\begin{equation}
	\begin{split}
		\langle \sigma_{ij}(\varepsilon) \rangle &= \frac{1}{V}\frac{\partial E}{\partial \varepsilon_{ij}}, \\
		&= C_{ijkl} \varepsilon_{kl} - \frac{1}{V} P_{ij}.
	\end{split}
	\label{eq:sigma_Pij}
\end{equation}
In the particular case where the periodicity vectors are kept fixed 
between the defective and pristine supercells ($\varepsilon=0$),
the elastic dipole is proportional to the residual stress weighted by the supercell volume:
\begin{equation}
	P_{ij} = -V \langle \sigma_{ij} \rangle. 
	\label{eq:Pij_from_sigma}
\end{equation}
This residual stress corresponds to the stress increase, after atomic relaxation, 
due to the introduction of the point-defect into the simulation box. 
When this equation is used to determine the elastic dipole in \abinitio calculations,
one should pay attention to the spurious stress which may exist 
in the equilibrium perfect supercell because of finite convergence criteria 
of such calculations.  This spurious stress has to be subtracted 
from the stress of the defective supercell, 
so the residual stress entering Eq. \ref{eq:Pij_from_sigma}
is only the stress increment associated with the introduction of the point-defect.

One can also consider the opposite situation where a homogeneous strain $\bar{\varepsilon}$
has been applied to cancel the residual stress.
The elastic dipole is then proportional to this homogeneous strain:
\begin{equation}
	P_{ij} = V C_{ijkl} \bar{\varepsilon}_{kl}.
	\label{eq:Pij_from_strain}
\end{equation}
One would nevertheless generally prefer working with fixed periodicity vectors 
($\varepsilon=0$) as $\sigma=0$ calculations necessitate an increased number 
of force calculations, as well as an increased precision for \abinitio calculations.
In the more general case where a homogeneous strain is applied 
and a residual stress is observed, the elastic dipole can still be derived 
from these two quantities using Eq. \eqref{eq:sigma_Pij}.

This definition of the elastic dipole from the residual stress (Eq. \ref{eq:Pij_from_sigma}),
or more generally from both the applied strain and the residual stress (Eq. \ref{eq:sigma_Pij}),
is to be related to the dipole tensor measurement first proposed by Gillan \cite{Gillan1981,Gillan1983},
where the elastic dipole is equal to the strain derivative of the formation energy, evaluated at zero strain. 
Instead of doing this derivative numerically, one can simply use the analytical derivative, 
\ie the stress on the simulation box, which is a standard output of any atomistic simulations code,
including \abinitio calculations.
This technique to extract elastic dipoles from atomistic simulations 
has been validated \cite{Subramanian2013,Garnier2014,Varvenne2017}, 
through successful comparisons of interaction energies between point-defects with external strain fields,
as given by direct atomistic simulations 
and as given by the elasticity theory predictions using the elastic dipole identified through Eq.~\eqref{eq:Pij_from_sigma}. 
The residual stress therefore leads to quantitative estimates of the elastic dipoles.

\subsubsection*{Definition from the displacement field}

The elastic dipole can also be obtained from the displacement field,
as proposed by Chen \etal~\cite{Chen2010a}.
Using the displacement field $\vec{u}^{\rm at}(\vec{R})$ obtained after relaxation
in atomistic simulations, a least-square fit of the displacement field $\vec{u}^{\rm el}(\vec{R})$
predicted by elasticity theory can be realized,
using the dipole components of the dipole as fit variables. 
A reasonable cost function for the least-square fit is
\begin{equation}
f(P_{ij}) =  \sum_{\substack{\vec{R} \\ \|\vec{R}\|>r_{\rm excl}}} \left\|R^2\left[\vec{u}^{\rm el}(\vec{R})-\vec{u}^{\rm at}(\vec{R})\right] \right\|^2 ,
\label{eq:cost_F}
\end{equation}
with $r_{\rm excl}$ the radius of a small zone around the point-defect, so as to exclude from the fit the atomic positions where elasticity does not hold. The $R^2$ factor accounts for the scaling of the displacement field with the distance to the point-defect, thus giving a similar weight to all atomic positions included into the fit.  
For atomistic simulations with periodic boundary conditions, 
one needs to superimpose the elastic displacements of the point-defect with its periodic images,
which can be done by simple summation, taking care of the conditional convergence of the corresponding sum \cite{Varvenne2017}.
With large simulation boxes ($\ge 1500$ atoms), the obtained elastic dipole components 
agree with the values deduced from the residual stress, and the choice of $r_{\rm excl}$ is not critical. 
The number of atomic positions included in the fit, and for which elasticity is valid, is sufficiently high to avoid issues arising from the defect core zone \cite{Varvenne2017}.
In contrast, for small simulation boxes of a few hundred atoms, \ie typical of \abinitio simulations, the obtained $P_{ij}$ values are highly sensitive  to $r_{\rm excl}$, and their convergence with  $r_{\rm excl}$ cannot be guaranteed.
This fit of the displacement field appears therefore impractical 
to obtain precise values of the elastic dipole in \abinitio calculations.

\subsubsection*{Definition from the Kanzaki forces}

\begin{figure}[!bt]
	\begin{center}
	\subfigure[unrelaxed vacancy]{\hspace*{5mm}\includegraphics[scale=0.75]{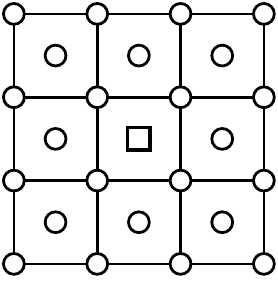}\hspace*{5mm}} 
	\subfigure[relaxed vacancy]{\hspace*{5mm}\includegraphics[scale=0.75]{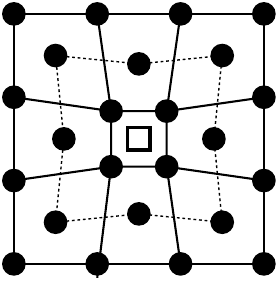}\hspace*{5mm}} \\
	\subfigure[$0^{\rm th}$ order approx.]{\hspace*{5mm}\includegraphics[scale=0.75]{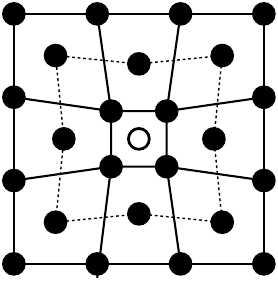}\hspace*{5mm}} 
	\subfigure[$1^{\rm st}$ order approx.]{\hspace*{5mm}\includegraphics[scale=0.75]{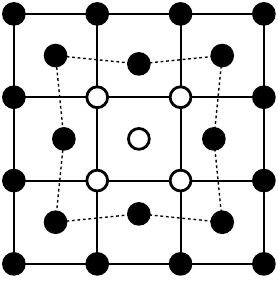}\hspace*{5mm}}
	\end{center}
	\caption{Procedure for the computation of the Kanzaki forces in the case of a vacancy. 
	The white spheres correspond to atoms at their perfect bulk positions, \ie before relaxation,
	the white square to the vacancy, and the black spheres to the atoms at their relaxed position around the defect.}
	\label{fig:scheme_kanzaki}
\end{figure}

The definition given in Eq. \eqref{eq:dipole} of the elastic dipole as the first moment of the point-force distribution
offers a third way to extract this elastic dipole from atomic simulations.
This corresponds to the Kanzaki force method 
\cite{Kanzaki1957,Faux1971,Tewary1973,Leibfried1978,Schober1980,Lidiard1981,Simonelli1994,Domain2004,Hayward2012}.
Kanzaki forces are defined as the forces which have to be applied to the atoms in the neighborhood of the point-defect 
to produce the same displacement field in the pristine crystal as in the defective supercell. 
Computation of these Kanzaki forces can be performed following the procedure given in Ref.~\cite{Simonelli1994}, which is illustrated for a vacancy in Fig.~\ref{fig:scheme_kanzaki}. 
Starting from the relaxed structure of the point-defect (Fig. \ref{fig:scheme_kanzaki}b),
the defect is restored in the simulation cell, 
\eg the suppressed atom is added back for the vacancy case (Fig.~\ref{fig:scheme_kanzaki}c).
A static force calculation is performed then and provides the opposite of the searched forces on all atoms in the obtained simulation cell. 
These atomic forces are used to compute the elastic dipole $P_{ij}=\sum_{q} F_j^q a_i^q$,
with $\vec{F}^q$ the opposite of the force acting on atom at $\vec{a}^q$, 
assuming the point-defect is located at the origin. 
The summation is usually restricted to atoms located inside a sphere of radius $r_{\rm \infty}$. 

As Kanzaki's technique is valid only in the harmonic approximation, one checks that the atomic forces entering the elastic dipole definition are in the harmonic regime by restoring larger and larger defect neighboring shells to their perfect bulk positions \cite{Simonelli1994} (Fig.~\ref{fig:scheme_kanzaki}c-d), computing the forces on the obtained restored structures, and then the elastic dipole. The case where $n$ defect neighbor shells are restored is referred to as the $n^{\rm th}$ order approximation.   
As the restored zone becomes larger, the atoms remaining at their relaxed positions are more likely to sit in an harmonic region. The convergence of the resulting elastic dipole components with respect to $n$ thus enables to evaluate the harmonicity aspect.   

\begin{figure}[!t]
	\begin{center}
	\includegraphics[scale=0.65,clip=true,trim=0mm 0mm 5mm 4mm]{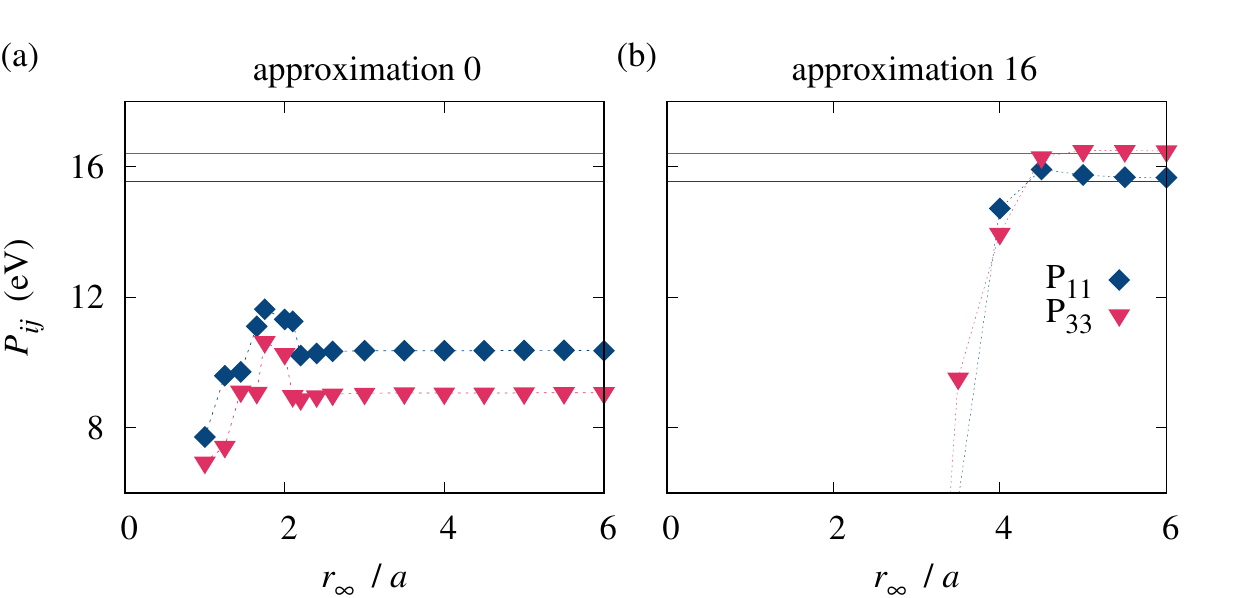}
	\end{center}
	\caption{Elastic dipole components of the SIA octahedral configuration in hcp Zr, 
	as a function of the cutoff radius $r_{\infty}$ of the force summation
	normalized by the lattice parameter $a$.
	Values are obtained by the Kanzaki's forces approach on a simulation box containing 12800 atoms, 
	restoring (a) only the point-defect, and (b) up to $16$ defect neighbor shells.
	The horizontal lines are the values deduced from the residual stress.
	Calculations have been performed with the EAM $\#3$ potential of Ref. \cite{Mendelev2007}
	(see Ref. \cite{Varvenne2017} for more details). 
	}
\label{fig:Pij_measurement_EAM}
\end{figure}

Fig. \ref{fig:Pij_measurement_EAM} provides the elastic dipole values as a function of the cutoff radius $r_{\infty}$,
for the octahedral configuration of the self-interstitial atom (SIA) in hcp Zr. 
Only the point-defect has been restored in Fig. \ref{fig:Pij_measurement_EAM}a (approximation 0),
whereas the restoration zone extends to the 16$^{\rm th}$ nearest-neighbors in Fig. \ref{fig:Pij_measurement_EAM}b.
Constant $P_{ij}$ values are reached for a cutoff radius $r_{\infty}\sim 2.5\,a$ and $\sim 4\,a$, respectively,
showing that the defect-induced forces are long-ranged
\cite{Hayward2012,Varvenne2017}. 
As a result, the supercell needs to be large enough to avoid convolution of the force field 
by periodic boundary conditions and a high precision on the atomic forces is required.
Comparing with the elastic dipole deduced from the residual stress, 
one cannot only restore the point-defect (approximation 0 in Fig. \ref{fig:Pij_measurement_EAM}a)
to obtain a quantitative estimate with the Kanzaki method.
A restoration zone extending at least to the 16$^{\rm th}$ nearest neighbors is necessary 
for this point-defect to obtain the correct elastic dipole.
As the anharmonic region depends on the defect and on the material, 
one cannot choose \textit{a priori} a radius for the restoration zone,
but one needs to check the convergence of the elastic dipole with the size of this restoration zone.

\subsubsection*{Discussion}

These three approaches lead to the same values of the elastic dipole when large enough supercells are used,
thus confirming the consistency of this elastic description of the point-defect.
This has been checked in Ref. \cite{Varvenne2017} for the vacancy and various configurations of the SIA
in hcp Zr. 
But for small simulation cells typical of \abinitio calculations, 
both the fit of the displacement field and the calculation from the Kanzaki forces 
are usually not precise enough because of the too large defect core region,
\ie the region which has to be excluded from the displacement fit
or the restoration zone for the Kanzaki forces.
This is penalizing for \abinitio calculations, even for point-defects as simple as the H solute or the vacancy in hcp Zr \cite{Varvenne2017,Nazarov2016}. 
Besides, the Kanzaki's technique requires additional calculations to obtain the defect-induced forces 
and to check that the forces entering the dipole definition are in the harmonic regime. 
As this restoration zone is extended, the defect-induced forces become smaller 
and the precision has to be increased.
The definition from the residual stress appears indeed as the only method 
leading to reliable $P_{ij}$ values within \abinitio simulations.
It is also easy to apply, as it does not require any post treatment nor additional calculations:
it only uses the homogeneous stress on the simulation box 
and the knowledge of the defect position is not needed.  

All these methods can be of course also used to determine the diaelastic polarizability.
One only needs to get the elastic dipole for various applied strains.
The linear equation \eqref{eq:polarizability} then leads
the stress-free elastic dipole $P^0_{ij}$ and the polarizability $\alpha_{ijkl}$.
The most convenient method remains a definition from the residual stress.
Considering the polarizability, Eq. \eqref{eq:sigma_Pij} now writes
\begin{equation}
	\langle \sigma_{ij}(\varepsilon) \rangle 
	= \left( C_{ijkl} - \frac{1}{V} \alpha_{ijkl} \right) \varepsilon_{kl} - \frac{1}{V} P_{ij},
	\label{eq:sigma_polarizability}
\end{equation}
thus showing that the polarizability is associated with a variation of the elastic constants
proportional to the point-defect volume fraction.
This linear variation of the elastic constants arising from the point-defect polarizability
has been characterized for vacancies and SIAs in face-centered cubic (fcc) copper \cite{Ackland1988},
or various solute atoms in body-centered cubic (bcc) iron \cite{Bialon2013,Fellinger2017}.

\begin{figure}[!bth]
	\centering
	\includegraphics[width=0.7\linewidth]{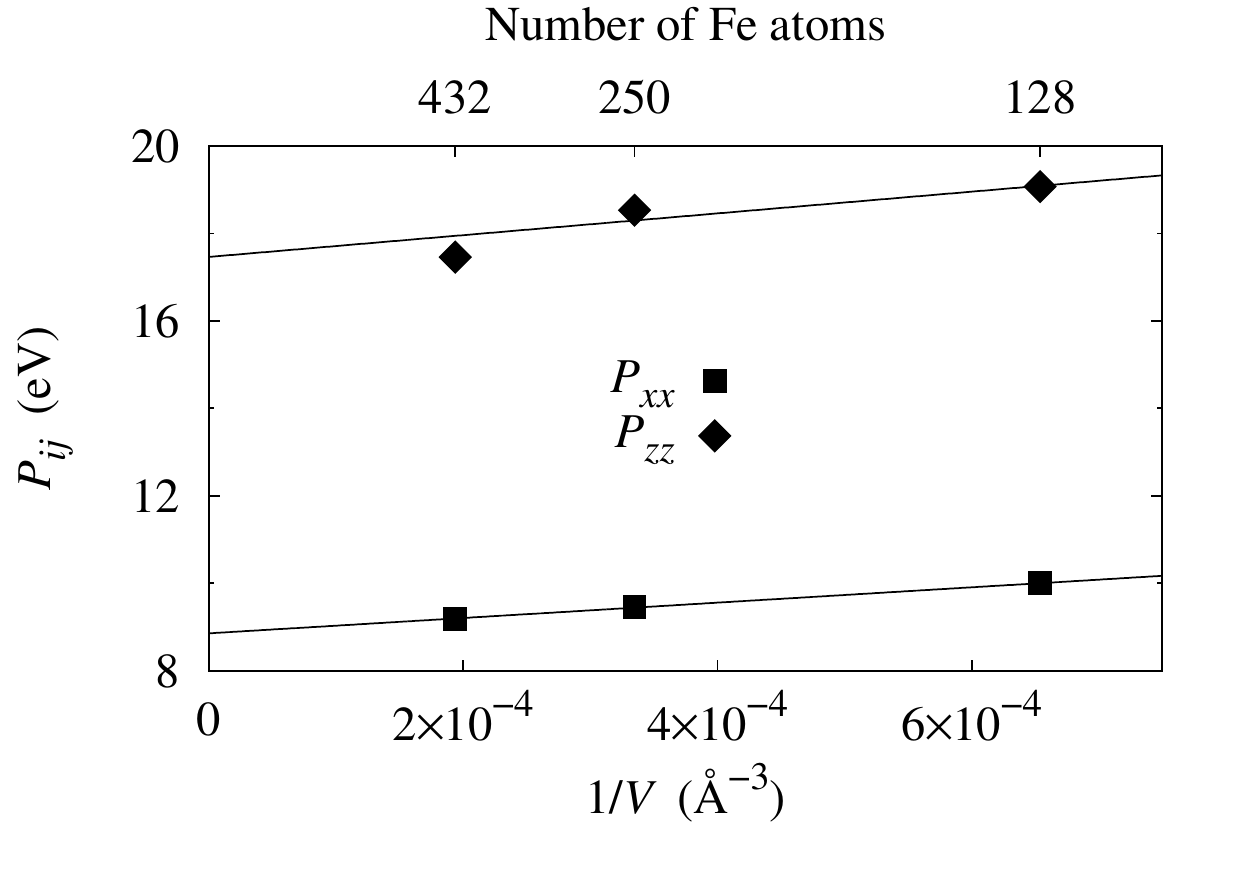}
	\caption{Elastic dipole of a C atom lying in a [001] octahedral interstitial site 
	in bcc Fe as a function of the inverse volume $V$ of the supercell.
	The elastic dipole has been deduced from the residual stress in \abinitio calculations
	(see Ref. \cite{Clouet2011b} for more details).}
	\label{fig:C_dipole}
\end{figure}

One consequence of the diaelastic polarizability is that the elastic dipole 
may depend on the size of the supercell with periodic boundary conditions. 
The strain at the point-defect position is indeed the superposition of the homogeneous strain $\varepsilon_{ij}$
and the strains created by the periodic images of the point-defect $\varepsilon_{ij}^{\rm p}$. 
In the $\varepsilon=0$ case for instance, the obtained elastic dipole is then
\begin{equation}
	P_{ij} = P^0_{ij} + \alpha_{ijkl} \varepsilon_{kl}^{\rm p}.
	\label{eq:dipole_PBC}
\end{equation}
As the strain created by a point-defect varies as the inverse of the cube of the separation distance
(Eq. \ref{eq:dipole_stress}), 
the last term in Eq. \eqref{eq:dipole_PBC} scales with the inverse of the supercell volume. 
Therefore, when homothetic supercells are used, 
one generally observes the following volume variation
\begin{equation*}
	P_{ij} = P^{0}_{ij} + \frac{\delta P_{ij}}{V},
\end{equation*}
which can be used to extrapolate the elastic dipole to an infinite volume,
\ie to the dilute limit \cite{Puchala2008,Clouet2011b,Varvenne2017}.
An example of this linear variation with the inverse volume 
is shown in Fig. \ref{fig:C_dipole} for an interstitial C atom in a bcc Fe matrix.

\subsection{From experiments}
\label{sec:para_exp}

From an experimental perspective, when trying to extract elastic dipole of point-defects, both the symmetry and the magnitude of the components of the elastic dipole tensor are \textit{a priori} unknown, and possibly also the number of defect-types into the material. 
We first restrict ourselves to the case where only one single type of point-defect with a known symmetry is present. 

If the point-defect has a lower symmetry than the host crystal, 
then it can adopt several variants which are equivalent by symmetry but possess  different orientations.
The energy of such a volume $V$ 
containing different variants of the point-defect and submitted to a homogeneous strain is
\begin{multline}
	E(\varepsilon)  =  E_0 + E^{\rm PD} + \frac{V}{2}C_{ijkl}\varepsilon_{ij}\varepsilon_{kl} 
	\\
	- V \sum_{\mu=1}^{n_{\rm v}}{ c_{\mu} P^{\mu}_{ij} }\varepsilon_{ij},
\end{multline}
with $n_{\rm v}$ the total number of different variants 
and $c_{\mu}$ the volume concentration of variant $\mu$.
This relation assumes that the different point-defects are not interacting, which is valid in the dilute limit. 
For zero stress conditions, as usually the case in experiments, the average strain 
induced by this assembly of point-defects is 
\begin{equation}
	\bar{\varepsilon}_{ij} = S_{ijkl} \sum_{\mu=1}^{n_{\rm v}}{ c_{\mu} P^{\mu}_{kl} },
	\label{eq:epsilon_Vegard}
\end{equation}
with $S_{ijkl}$ the inverse of the elastic constants $C_{ijkl}$.
This linear relation between the strain and the point-defect concentrations corresponds to a Vegard's law 
and allows for many connections with experiments. 
It generalizes Eq. \ref{eq:Pij_from_strain} to the case of a volume 
containing a population of the same point-defect with different variants.
As mentioned in \S\ref{sec:dipole_model}, point-defects in experiments are sometimes 
rather characterized by their $\lambda$-tensor \cite{Nowick1972}. 
Combining the definition of this $\lambda$-tensor (Eq. \ref{eq:lambda_PD}) with Eq. \eqref{eq:epsilon_Vegard},
one shows the equivalence of both definitions:
\begin{equation*}
	\lambda_{ij}^{\mu} = \frac{1}{\Omega_{\rm at}} \, S_{ijkl} \, P^{\mu}_{kl},
\end{equation*}
or equivalently Eq. \eqref{eq:dipole_lamba}.

When the point-defect has only one variant or when only one variant is selected by breaking the symmetry 
-- through either a phase transformation (\eg martensitic  \cite{Roberts1953,Cheng1990})
or the interaction with an applied strain field for instance --
the variations of the material lattice constants 
with the defect concentration follow the defect symmetry.
If the point-defect concentration is known, the elastic dipole components are therefore fully accessible
by measuring lattice parameter variations, \eg by dilatometry or X-ray diffraction using the Bragg reflections. 

On the other hand, for a completely disordered solid solution of point-defects 
with various variants ($n_{\rm v}>1$), 
the average distortion induced by the point-defect population does not modify the parent crystal symmetry \cite{Nowick1972}. 
Each variant is equiprobable, \ie $c_{\mu}=c_0/n_{\rm v}$ with $c_0$ the nominal point-defect concentration. 
The stress-free strain induced by the point-defect (Eq. \ref{eq:epsilon_Vegard}) 
thus becomes
\begin{equation*}
	\bar{\varepsilon}_{ij} = c_0 \, S_{ijkl} \, \langle P_{kl} \rangle
	\ \textrm{ with }\ 
	\langle P_{kl} \rangle = \frac{1}{n_{\rm v}} \sum_{\mu=1}^{n_{\rm v}}{ P^{\mu}_{kl} }.
\end{equation*}
Measurements of the lattice parameter variations with the total defect concentration 
give thus access only to some sets of combinations of the $P_{ij}$ components. 
For instance, if we consider a point-defect in a cubic crystal,
like a C solute in an octahedral site of a bcc Fe crystal, 
one obtains the following variation of the lattice parameter with the solute concentration
\begin{equation}
	a(c_0) = a_0 \left( 1 + \frac{\Tr{(P)}}{3 \left(C_{11}+2C_{12}\right)} \, c_0 \right),
	\label{eq:lattice_change_cubic}
\end{equation} 
with $C_{11}$ and $C_{12}$ the elastic constants in Voigt notation. 
This variation can again be characterized using dilatometry or X-ray diffraction. 
But knowing $\Tr{(P)}$ is not sufficient for a point-defect with a lower symmetry than the cubic symmetry of the crystal, as
the elastic dipole has several independent components (two for the C solute atom in bcc Fe).
Additional information is therefore needed to fully characterize the point-defect. 

For those defects having a lower symmetry than their parent crystal, the anelastic relaxation experiments may provide such supplementary data \cite{Nowick1972, Nowick1963}. By applying an appropriate stress, a splitting of the point-defect energy levels occurs, and a redistribution of the defect populations is operated. The relaxation of the compliance moduli then gives access to other combinations of the elastic dipole components. 
Not all of the relaxations are allowed by symmetry, 
as illustrated for the C solute in bcc Fe, where only the quantity $|P_{11}-P_{33}|$ is accessible \cite{Swartz1968}. 
The number of parameters accessible from anelastic measurements is lower 
than the independent components of the defect elastic dipole.
This technique must then be used in combination with other measurements,
like the variations of the lattice parameter.   

Alternatively, a useful technique working with a random defect distribution is the diffuse Huang scattering. The diffuse scattering of X-rays near Bragg reflections  \cite{Trinkaus1972,Bender1983,Michelitsch1996} reflects the distortion scattering caused by the long-range part of the defect-induced displacement field.
It thus provides information about the strength of the point-defect elastic dipole.   
The scattered intensity is proportional -- in the dilute limit -- to the defect concentration and to a linear combination of quadratic expressions of the elastic dipole components. The coefficients of this combination are functions of the crystal elastic constants and of the scattering vector in the vicinity of a given reciprocal lattice vector. Therefore, by an appropriate choice of the relative scattering direction, the quadratic expressions can be determined separately. 

Except for simple point-defects like a substitutional solute atom or a single vacancy, 
the defect symmetry may be unknown.
Both anelastic relaxation and Huang scattering experiments provide important information
for the determination of the defect symmetry. 
The presence of relaxation peaks in anelasticity 
is a direct consequence of the defect symmetry \cite{Nowick1972,Nowick1963}.
Within Huang scattering experiments, information about the defect symmetry is obtained either by the analysis of the morphology of iso-intensity curves or through an appropriate choice of scattering directions to measure the Huang intensity. 

To conclude, when extracting elastic dipoles from experiments, 
one must usually rely on a combination of several experimental techniques to obtain all components.

\section{Some applications}
\label{sec:examples}

\subsection{Solute interaction with a dislocation}

\begin{figure}[!bt]
	\centering
	\subfigure[Screw dislocation ($h=4d_{110}\simeq8.1$\,\AA)]{
		\includegraphics[width=0.8\linewidth]{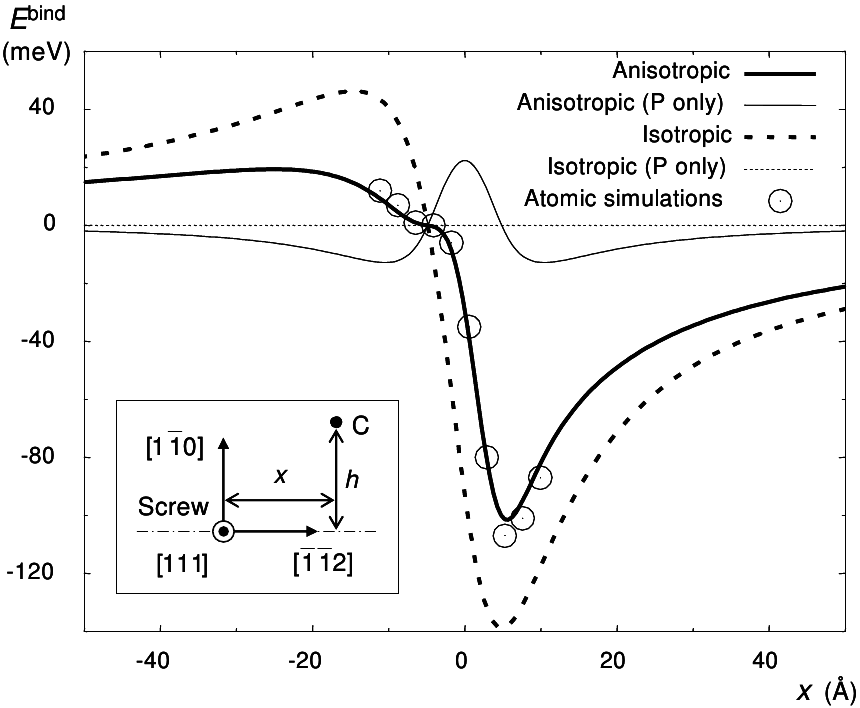}}
	\subfigure[Edge dislocation ($h=-9d_{110}\simeq-18.2$\,\AA)]{
		\includegraphics[width=0.8\linewidth]{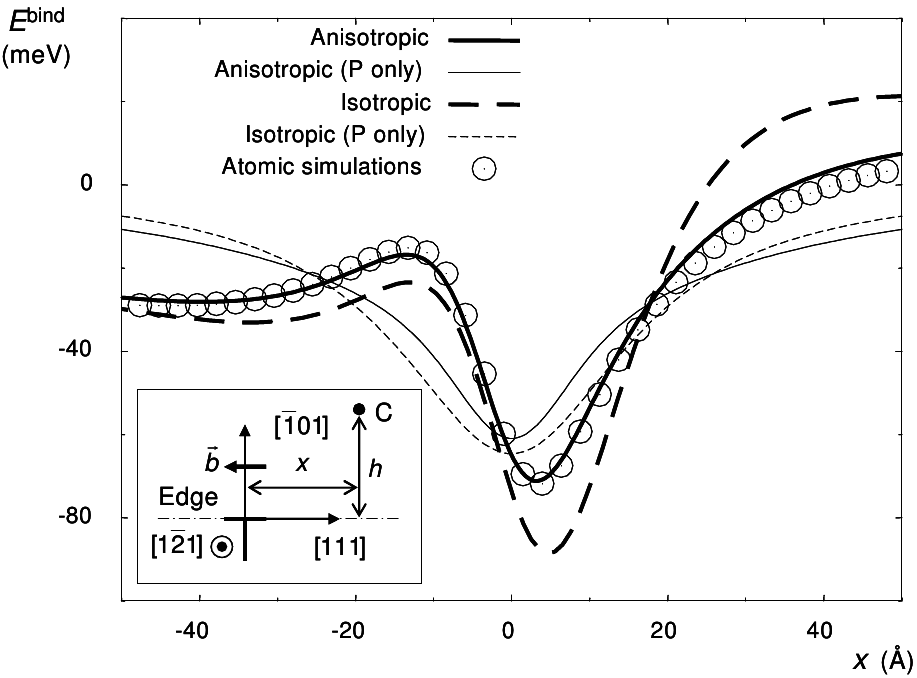}}
	\caption{Binding energy $E^{\rm bind} = -E^{\rm int}$
	between a screw or an edge dislocation and a C atom  in bcc iron
	for different positions $x$ of the dislocation in its glide plane.
	The C atom is lying in a [100] octahedral interstitial site at a fixed distance $h$ 
	of the dislocation glide plane.
	Symbols correspond to atomistic simulations and lines to elasticity theory, 
	considering all components of the stress created by the dislocation or only the pressure,
	and using isotropic or anisotropic elasticity.}
	\label{fig:dislo_Fe_C}
\end{figure}

This elastic modeling can be used for instance to describe the interaction 
of a point-defect with other structural defects. 
To illustrate, and also validate, this approach, we consider a C interstitial atom
interacting with a dislocation in a bcc iron matrix. 
This interstitial atom occupies the octahedral sites of the bcc lattice. 
As these sites have a tetragonal symmetry, the elastic dipole $P_{ij}$ of the C atoms has two independent
components and gives thus rise to both a size and a shape interaction.
The interaction energy of the C atom with a dislocation 
is given by Eq. \eqref{eq:dipole_Einter} where the external strain $\varepsilon^{\rm ext}_{ij}$
is the strain created by the dislocation at the position of the C atom.
This has been compared in Ref. \cite{Clouet2008}
to direct results of atomistic simulations, using for the C elastic dipole and for the elastic constants
the values given by the empirical potential used for the atomistic simulations. 
Results show that elastic theory leads to a quantitative prediction when all ingredients are included 
in the elastic model, \ie when elastic anisotropy is taken into account to calculate the strain field created
by the dislocation and when both the dilatation and the tetragonal distortion 
induced by the C atom are considered (Fig. \ref{fig:dislo_Fe_C}).
The agreement between both techniques is perfect 
except when the C atom is in the dislocation core. 
With isotropic elasticity, the agreement with atomistic simulations is only qualitative, 
and when the shape interaction is not considered, \ie when the C atom is modeled as a simple dilatation center 
($P_{ij}=P\,\delta_{ij}$), elastic theory fails to predict this interaction (Fig. \ref{fig:dislo_Fe_C}).
The same comparison between atomistic simulations and elasticity theory has been performed 
for a vacancy and a SIA interacting with a screw dislocation still in bcc iron \cite{Hayward2012}. 
The agreement was not as good as for the C atom. But in this work, the elastic dipoles of the point-defects
were obtained from the Kanzaki forces, using the $0^{\rm th}$ order approximation, which is usually not as precise as the definition from the stress 
(\cf \S\,\ref{sec:para_atom}) and may explain some of the discrepancies. 

One can also use elasticity theory to predict how 
the migration barriers of the point-defect are modified by a strain field.
The migration energy is the energy difference between the saddle point and the stable position. 
Its dependence with an applied strain field $\varepsilon(\vec{r})$ is thus described by
\begin{equation}
	E^{\rm m}[\varepsilon] = E^{\rm m}_0 
	+ P_{ij}^{\rm ini} \varepsilon_{ij}(\vec{r}_{\rm ini}) 
	- P_{ij}^{\rm sad} \varepsilon_{ij}(\vec{r}_{\rm sad}) ,
	\label{eq:Emig_strain}
\end{equation}
where $P_{ij}^{\rm ini}$ and $P_{ij}^{\rm sad}$ are the elastic dipoles of the point-defect
respectively at its initial stable position $\vec{r}_{\rm ini}$
and at the saddle point $\vec{r}_{\rm sad}$,
and $E^{\rm m}_0$ is the migration energy without elastic interaction.
Still for a C atom interacting with a dislocation in a bcc Fe matrix, 
comparison of this expression with results of direct atomistic simulations
show a good agreement \cite{Veiga2011}, as soon as the C atom is far enough 
from the dislocation core.
Similar conclusions, on the validity of equation \eqref{eq:Emig_strain} 
to describe the variation of the solute migration energy
with an applied strain, have been reached
for a SIA diffusing in bcc Fe \cite{Chen2010a},
a vacancy in hcp zirconium \cite{Subramanian2013}
or a Si impurity in fcc nickel \cite{Garnier2014}.

\subsection{Elastodiffusion}

This simple model predicting the variation of the migration energy
with an applied strain field (Eq. \ref{eq:Emig_strain}) can be used to study elastodiffusion. 
Elastodiffusion refers to the 
diffusion variations 
induced by an elastic field \cite{Dederichs1978}, either externally applied or internal through the presence of structural defects. Important implications exist for materials, such as transport and segregation of point-defects to dislocations
leading to the formation of Cottrell atmospheres \cite{Cottrell1949},
irradiation creep \cite{Woo1984}, 
or anisotropic diffusion of dopants in semiconductor thin films \cite{Aziz1997,Daw2001}.

At the atomic scale, solid state diffusion occurs through the succession of thermally activated atomic jumps from stable to other stable positions,
with atoms jumping either on vacancy sites or on interstitial sites of the host lattice. 
Within transition state theory \cite{Vineyard1957}, the frequency of such a transition is given by
\begin{equation}
	\Gamma_{\alpha} = \nu^0_{\alpha} \exp{ \left( - E^{\rm m}_{\alpha}\,/\, kT \right)},
	\label{eq:transition_rate}
\end{equation}
where $\nu^0_{\alpha}$ is the attempt frequency for the transition $\alpha$ 
and $E^{\rm m}_{\alpha}$ is the migration energy.

Considering the effect of a small strain field on this bulk system, the diffusion network and the site topology will not be modified.
On the other hand, the presence of this small strain field modifies the migration energies 
and the attempt frequencies.   
As shown in the previous section, the elastic dipole description of the point-defect 
can predict the modification of the stable and saddle point energies, and thus of the migration energy
(Eq. \ref{eq:Emig_strain}).
Ignoring the strain effect on attempt frequencies, 
the incorporation of the modified energy barriers into stochastic simulations
like atomistic or object kinetic Monte Carlo (OKMC) methods
enables to characterize the point-defect elastodiffusion effect.
This approach has been used, for instance, to study the directional diffusion of point-defects
in the heterogeneous strain field of a dislocation, 
corresponding to a biased random walk  \cite{Veiga2010,Veiga2011,Subramanian2013}. 

Diffusion in a continuous solid body is characterized by the diffusion tensor $D_{ij}$
which expresses the proportionality between the diffusion flux and the concentration gradient (Fick's law).
The effect of an applied strain is then described by the elastodiffusion fourth-rank tensor $d_{ijkl}$ 
\cite{Dederichs1978}, 
which gives the linear dependence of the diffusion tensor with the strain:
\begin{equation}
D_{ij} = D_{ij}^0 + d_{ijkl} \, \varepsilon_{kl}.
\label{eq:d_ijkl}
\end{equation}
This elastodiffusion tensor obeys the minor symmetries 
$d_{ijkl}=d_{jikl}=d_{ijlk}$, because of the symmetry of the diffusion and deformation tensors,
and also the crystal symmetries.
Starting from the atomistic events as defined by their transition frequencies 
(Eq. \ref{eq:transition_rate}),
the diffusion coefficient, and its variation under an applied strain, 
can be evaluated from the long time evolution
of the point-defect trajectories in stochastic simulations \cite{Goyal2015}. 
Alternatively, analytical approaches can be developed to provide expressions
\cite{Howard1964,Allnatt1993}.
The elastodiffusion can thus be computed by a perturbative approach, 
starting from the analytical expression of the diffusion tensor \cite{Dederichs1978,Trinkle2016}.
This results in two different contributions: 
a geometrical contribution caused by the overall change of the jump vectors
and a contribution due to the change in energy barriers as described by Eq. \eqref{eq:Emig_strain}. 
This last contribution is thus a function of the elastic dipoles 
at the saddle point and stable positions.
It is found to have an important magnitude in various systems \cite{Dederichs1978,Trinkle2016},
being for instance notably predominant for interstitial impurities in hcp Mg  \cite{Agarwal2016}.
It is temperature-dependent, 
sometimes leading to complex variations with non-monotonic variations
and also sign changes for some of its components \cite{Agarwal2016}.
As noted by Dederichs and Schroeder \cite{Dederichs1978}, the elastic dipole at the saddle point 
completely determines the stress-induced diffusion anisotropy
in cubic crystals.
Experimental measurement of the elastodiffusion tensor components can therefore provide useful information about the saddle point configurations. 
 
Both approaches, relying either on stochastic simulations or analytical models, 
are now usually informed with \abinitio computed formation and migration energies, and attempt frequencies. 
The elastic modeling of a point-defect through its elastic dipole offers thus a convenient way 
to transfer the information about the effects of an applied strain, as obtained from atomistic simulations, 
to the diffusion framework.

\subsection{Bias calculations}

Point-defect diffusion and absorption by elements of the microstructure such as dislocations, 
cavities, grain boundaries and precipitates play an important role in the macroscopic evolution of materials.
It is especially true under irradiation, since in this case not only vacancies 
but also self-interstitial atoms (SIAs) migrate to these sinks. 
Owing to their large dipole tensor components, SIAs generally interact more than vacancies with the stress fields generated by sinks. This leads to a difference in point-defect fluxes to a given sink known as the ``absorption bias''. For example, in the ``dislocation bias model''~\cite{Brailsford1972}, which is one of the most popular models to explain irradiation void swelling, 
dislocations are known as biased sinks: they absorb more interstitials than vacancies. 
Voids, which produce shorter range stress fields, are considered as neutral sinks, meaning that their absorption bias is zero. Since SIAs and vacancies are produced in the same quantity, the preferential absorption of SIAs by dislocations leads to a net flux of vacancies to voids and thus to void growth. Similar explanations based on absorption biases have been given to rationalize irradiation creep~\cite{Heald1974} and irradiation growth in hexagonal materials~\cite{Rouchette2014a}. In order to predict the kinetics of such phenomena, a precise evaluation of absorption biases is necessary.

Following the rate theory formalism \cite{Brailsford1972}, 
the absorption bias of a given sink can be written as the relative difference of sink strengths 
for interstitials ($k_i^2$) and vacancies ($k_v^2$)~\cite{Heald1975a}. 
The strength of a sink for a point-defect $\theta$ ($\theta = i, v$) is related to the loss rate
$\phi_\theta$ through
\begin{equation}
  \label{eq-flux-sink-strength}
  \phi_{\theta} = k_{\theta}^2 D_{\theta} c_{\theta},
\end{equation}
where $D_\theta$ is the diffusion coefficient free of elastic interactions 
and $c_{\theta}$ is the volume concentration of $\theta$.

The sink strength can be calculated with different methods, for example by solving the diffusion equation around the sink~\cite{Brailsford1972,Dederichs1978} or an associated phase field model~\cite{Rouchette2014}, or by performing object kinetic Monte Carlo simulations (OKMC)~\cite{Heinisch2000,Malerba2007}. It should be noted that analytical solution of the diffusion equation is limited to a few cases and often requires the defect properties or the stress field to be simplified~\cite{Schroeder1975,Woo1981,Skinner1984}, so in general numerical simulations are necessary~\cite{Woo1979a,Bullough1981,Dubinko2005,Jourdan2015}. In the following we consider the OKMC approach, due to its simplicity and its flexibility to introduce complex diffusion mechanisms and the effect of stress fields~\cite{Sivak2011,Subramanian2013,Vattre2016}.

In OKMC simulations of sink strengths, a sink is introduced in a simulation box where periodic boundary conditions are used and point-defects are generated at a given rate $K$. They diffuse in the box by successive atomic jumps until they are absorbed by the sink. For each defect in the simulation box, the jump frequencies of all jumps from the current stable state to the possible final states are calculated and the next event is chosen according to the standard residence time algorithm~\cite{Gillespie1976,Bortz1975}. 
The jump frequency of event $\alpha$ is given by Eq. \eqref{eq:transition_rate},
considering the strain dependence of the migration energy through Eq. \eqref{eq:Emig_strain}.

The sink strength is deduced from the average number of defects in the box $\overline{N}_{\theta}$ at steady state by the following equation~\cite{Vattre2016}:
\begin{equation}
  \label{eq-sink-strength-from-C}
  k_{\theta}^2 = \frac{K}{D_{\theta}\overline{N}_{\theta}},
\end{equation}
from which the bias is deduced:
\begin{equation}
  \label{eq-bias-definition}
  B = \frac{k_i^2-k_v^2}{k_i^2}.
\end{equation}

Another method is often used for the calculation of sink strengths with OKMC~\cite{Heinisch2000,Malerba2007}. For each defect, the number of jumps it performs before it is absorbed by the sink is registered. The sink strength is then deduced from the average number of jumps. Although this method is equivalent to the method based on the average concentration in the non-interacting case, it is no more valid if elastic interactions are included. In this case the average time before absorption should be measured instead of the average number of jumps, since jump frequencies now depend on the location of the defect and are usually higher. Therefore, applying this method in the interacting case often leads to an underestimation of sink strengths.

As an illustration, we consider the study published in Ref. \cite{Vattre2016},
where sink strengths of semi-coherent interfaces have been calculated with OKMC, 
taking into account the effect of the strain field generated by the interfaces.
The strain is the sum of the coherency strain and of the strain due to interface dislocations. 
It has been calculated by a semi-analytical method within the framework of anisotropic elasticity
\cite{Vattre2013,Vattre2015,Vattre2016}. 
We consider the case of a twist grain boundary in Ag, 
which produces a purely deviatoric strain field.
Two grain boundaries distant from each other by $d$ are introduced in the box
and periodic boundary conditions are applied.

Dipole tensors of vacancies and SIAs in Ag have been computed by DFT for both stable and saddle positions~\cite{Vattre2016}, 
using the residual stress definition (Eq. \ref{eq:Pij_from_sigma}). 
At the ground state, the elastic dipole of the vacancy is isotropic and the one of the SIA almost isotropic. 
On the other hand, the elastic dipole tensors have a significant deviatoric component 
for both point-defects at their saddle point. 

\begin{figure}[!bt]
  \centering
  \includegraphics[width=\linewidth]{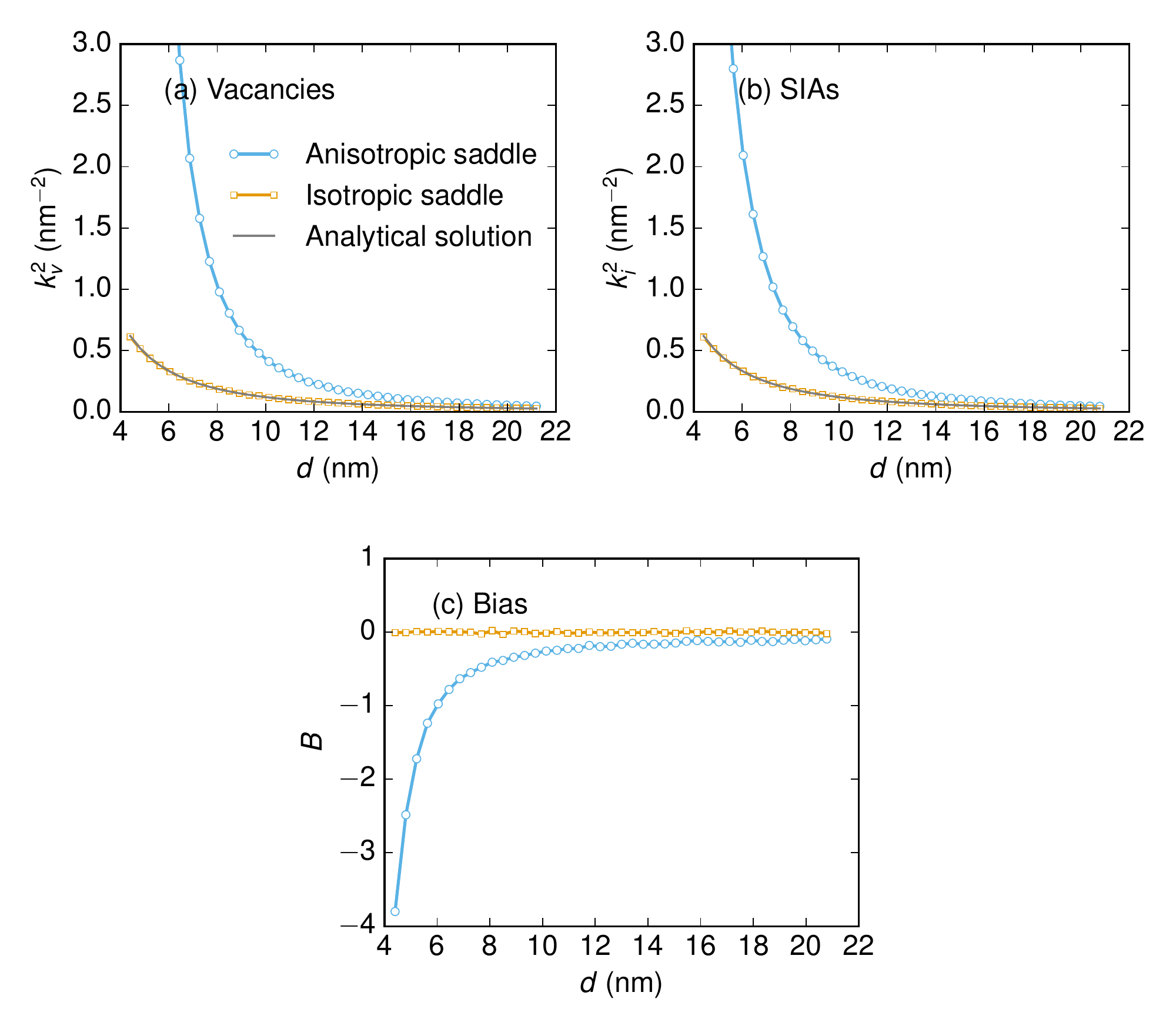}
  \caption{Sink strengths of a twist grain boundary ($\theta = 7.5$\textdegree) for (a) vacancies and (b) SIAs, 
  and (c) absorption bias, as a function of the layer thickness $d$.
  (see Ref. \cite{Vattre2016} for more details).}
  \label{fig-sink-strengths-and-bias}
\end{figure}

Sink strengths of  the twist grain boundary are shown in Fig.~\ref{fig-sink-strengths-and-bias}-(a,b) 
as a function of the layer thickness $d$ and compared to the analytical result with no elastic interactions
$k^2 = 12/d^2$. 
Sink strengths for both vacancies and SIAs are significantly increased 
when elastic interactions are included and when anisotropy at saddle point is taken into account, 
especially for thinner layers.   
However, if the saddle point is considered isotropic, the non-interacting case is recovered.
This is due to the deviatoric character of the strain field:
since the dipole tensor of the vacancy in its ground state is purely hydrostatic, 
the interaction energy of a vacancy with the strain field is zero 
and there is no thermodynamic driving force for the absorption of the vacancy. 
A similar result is obtained for SIAs,  because of their almost purely hydrostatic dipole for their ground state. 
Fig.~\ref{fig-sink-strengths-and-bias}c shows the evolution of the bias. 
For this interface, saddle point anisotropy leads to a negative bias, 
meaning that vacancies tend to be more absorbed than interstitials. 

This approach has also been recently used for the calculation of the sink strength of straight dislocations
and cavities in aluminum \cite{Carpentier2017}.
In both cases, saddle point anisotropy appears to have a significant influence on the sink strengths.
This confirms analytical results obtained with various levels of approximation
\cite{Skinner1984,Borodin1993,Borodin1994}.

\subsection{Isolated defect in atomistic simulations}
\label{sec:elast_corr}

The elastic modeling of point-defects is also useful in the context of atomistic simulations.
Such simulations, in particular \abinitio calculations, are now unavoidable to obtain the point-defects energetics, 
like their formation and migration energies \cite{Freysoldt2014}.
However, an ongoing issue is their difficulty to obtain the properties of isolated defects.
One can use atomistic simulations with controlled surface to model an isolated point-defect
\cite{Sinclair1978,Rao1998,Liu2007,Zhang2013,Huber2016}, 
but then, the excess energy associated with the point-defect 
could be exactly set apart from the one of the external surfaces or interfaces
only for interatomic potentials with a cutoff interaction radius, 
corresponding to short-range empirical potentials
like EAM. 
For more complex potentials or for \abinitio calculations, 
the absence of any interaction cutoff prevents an unambiguous definition of the point-defect energy.
A supercell approach relying on periodic boundary conditions is therefore usually preferred.
The combined effect of periodic boundary conditions 
and of the limited size of such calculations, for numerical cost reasons, 
makes the computed properties difficult to converge for defects inducing long-range effects. 
This problem is well-known in the context of charged point-defects, 
where long-range Coulombian interactions exist between the defect and its periodic images
and for which corrective schemes have been developed \cite{Leslie1985,Makov1995,Taylor2011}. 
For neutral defects, interactions between periodic images also exist. 
These interactions are of elastic origin and decay like the inverse cube of the separation distance.
Consequently, the computed excess energies are those of a periodic array of interacting  point-defects, 
and converge with the inverse of the supercell volume to the energy of the isolated defect.
This can be penalizing for defects inducing large distortions, like SIAs or clusters,
or for atomic calculations where only small supercells are reachable.
The elastic description of a point-defect allows calculating this spurious elastic interaction associated with periodic boundary conditions
to obtain the energy properties of the isolated point-defect \cite{Varvenne2013}.

After atomic relaxation, the excess energy of a supercell containing one point-defect 
is given by: 
\begin{equation}
	\label{eq:E_DP}
	E^{\rm PD}_{\rm PBC}(\bar{\varepsilon}=0) = E_{\infty}^{\rm PD} + \frac{1}{2} E_{\rm PBC}^{\rm int},
\end{equation}
where $E_{\infty}^{\rm PD}$ is the excess energy of the isolated defect 
and $E_{\rm PBC}^{\rm int}$ is the interaction energy of the defect with its periodic images.
The factor $1/2$ arises because half of the interaction is devoted to the defect itself
and the other goes to its periodic images.  
Continuous linear elasticity theory can be used to evaluate this elastic interaction.
If the point-defect is characterized by the elastic dipole $P_{ij}$,
following Eq. \ref{eq:dipole_Einter},
this interaction energy is given by
\begin{equation}
	E_{\rm PBC}^{\rm int} = - P_{ij} \, \varepsilon^{\rm PBC}_{ij},
	\label{eq:Epint} 
\end{equation}
with $\varepsilon^{\rm PBC}_{ij}$ the strain created by the defect periodic images.
It can be obtained by direct summation
\begin{equation}
	\varepsilon^{\rm PBC}_{ij} = -{\sum_{n,m,p}}'G_{ik,jl}(n\vec{a}_1+m\vec{a}_2+p\vec{a}_3 ) \, P_{kl}.
	\label{eq:eps_p}
\end{equation}
with  $\vec{a}_1$, $\vec{a}_2$ and $\vec{a}_3$ the periodicity vectors of the supercell.
The prime sign indicates that the diverging term ($n=m=p=0$) has been excluded from the sum.
As the second derivative of the Green's function $G_{ik,jl}(\vec{r})$ is decaying like $1/r^3$, 
this sum is only conditionally convergent.
It can be regularized following the numerical scheme proposed by Cai \cite{Cai2003}.

After computing the point-defect energy with an atomistic simulation code,
this energy can be corrected by subtracting the interaction energy with the periodic images (Eq. \ref{eq:E_DP})
to obtain the properties of the isolated defect. 
This interaction energy is computed from the elastic constants of the perfect crystal, 
which are needed to evaluate the Green's function and its derivative (\cf \S\,\ref{sec:elast_Hooke}),
and from the residual stress of the defective supercell to determine the point-defect elastic dipole (\cf \S\,\ref{sec:para_atom}).
This is therefore a simple post-treatment, which does not involve any fitting procedure
and which can be performed using the \textsc{Aneto} program provided as supplemental material of Ref. \cite{Varvenne2013}.

We have assumed in Eq. \eqref{eq:E_DP} that the supercell containing the point-defect 
has the same periodicity vector than the perfect supercell, 
\ie the applied homogenous strain $\bar{\varepsilon}$ is null. 
This corresponds to the easiest boundary conditions in atomistic simulations of point-defects. 
But sometimes, one prefers to relax also the periodicity vectors 
to nullify the stress  in the supercell.
Both these $\bar{\varepsilon}=0$ and $\sigma=0$ conditions converge to the same energy $E^{\rm PD}_{\infty}$
in the thermodynamic limit but different energies are obtained for too small supercells. 
The elastic model can be further developed to rationalize this difference \cite{Puchala2008,Varvenne2013}. 
For $\sigma=0$ conditions, a strain $\bar{\varepsilon}$ is
applied to the defective supercell to nullify its stress.
Eq. \eqref{eq:E_DP} therefore needs to be complemented with the energy contribution of this deformation
\begin{equation*}
	\Delta E(\bar{\varepsilon}) = 
		\frac{V}{2}C_{ijkl}\bar{\varepsilon}_{ij}\bar{\varepsilon}_{kl}
		- P_{ij} \bar{\varepsilon}_{ij}.
\end{equation*}
This applied strain $\bar{\varepsilon}$ in zero stress calculations is linked to the elastic dipole 
by Eq. \eqref{eq:Pij_from_strain}. 
The excess energy of the supercell containing one point-defect is thus now given by
\begin{equation}
	\label{eq:E_DP_sig0}
	\begin{split}
	E^{\rm PD}_{\rm PBC}(\sigma=0) &= E_{\infty}^{\rm PD} + \frac{1}{2} E_{\rm PBC}^{\rm int} - \frac{1}{2V}S_{ijkl}P_{ij}P_{kl} \\
	& = E^{\rm PD}_{\rm PBC}(\bar{\varepsilon}=0) - \frac{1}{2V}S_{ijkl}P_{ij}P_{kl},
	\end{split}
\end{equation}
where the elastic compliances of the bulk material $S_{ijkl}$ are the inverse tensor of the elastic constants $C_{ijkl}$. 
This equation shows that $\bar{\varepsilon}=0$ and $\sigma=0$ conditions lead to point-defect excess energies 
differing by a factor proportional to the inverse of the supercell volume
and to the square of the elastic dipole. 
This difference will be therefore important for small supercells and/or point-defects 
inducing an important perturbation of the host lattice. 
But once corrected through Eqs. \eqref{eq:E_DP} or \eqref{eq:E_DP_sig0}, 
both approaches should lead to the same value.
$\sigma=0$ calculations appear therefore unnecessary.

\begin{figure}[!ht]
	\includegraphics[scale=0.8]{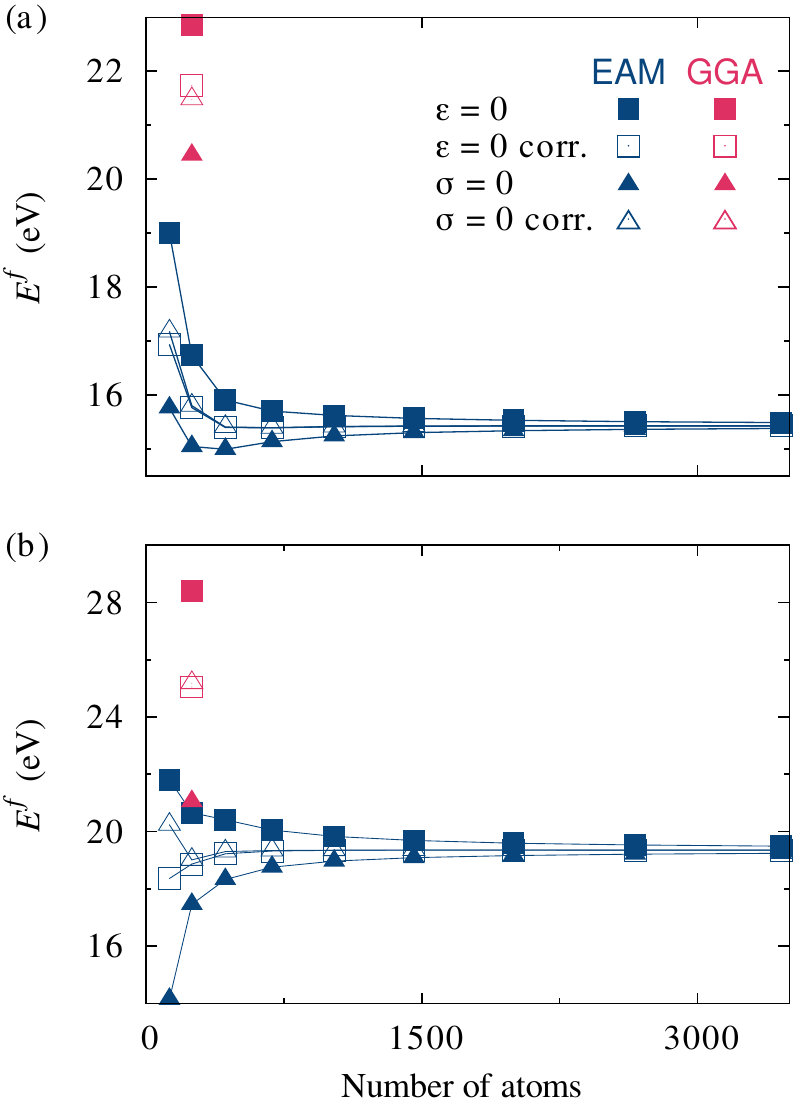}
\caption{Formation energy of a SIA cluster containing eight interstitials in bcc iron
calculated for fixed periodicity vectors ($\bar{\varepsilon} = 0$) or at zero stress ($\sigma=0$)
for different sizes of the simulation cell: 
(a) C15 aggregate and (b) parallel-dumbell configuration with a $\langle111\rangle$ orientation.
Atomistic simulations are performed either with the M07 empirical potential \cite{Marinica2012} (EAM)
or with \abinitio calculations (GGA).
Filled symbols refer to uncorrected results and open symbols to the results corrected by the elastic model 
(see Ref. \cite{Varvenne2013} for more details).}
\label{fig:Ef_8sia111c15}
\end{figure}

We illustrate the usefulness of this elastic post-treatment 
on an atomistic study of SIA clusters in bcc iron.
These clusters appear under irradiation and can adopt different morphologies \cite{Marinica2012}. 
In particular, some clusters can have a 3D structure with an underlying crystal symmetry corresponding to the C15 Laves' phase, 
and others have a planar structure corresponding to dislocation loop clusters with $1/2\,\langle111\rangle$ Burgers vectors. 

The formation energies of two different configurations of a cluster containing 8 SIAs, 
a C15 aggregate and a planar aggregate of parallel-dumbells with a $\langle 111 \rangle$ orientation,
are shown in Fig.~\ref{fig:Ef_8sia111c15} for different supercell sizes. 
They have been first calculated with an empirical EAM potential \cite{Marinica2012}: 
with fixed periodicity vectors ($\bar{\varepsilon}=0$), 
one needs at least $2000$ atoms for the C15 aggregate 
and $4000$ atoms for the $\langle 111 \rangle$ planar configuration 
to get a formation energy converged to a precision better than $0.1$\,eV. 
The convergence is slightly faster for zero stress calculations ($\sigma=0$) 
in the case of the C15 aggregate (Fig. \ref{fig:Ef_8sia111c15}a), 
but the opposite is true in the case of the $\langle 111 \rangle$ planar configuration (Fig. \ref{fig:Ef_8sia111c15}b).
When we add the elastic correction, the convergence is improved for both cluster configurations.
The corrected $\bar{\varepsilon}=0$ and $\sigma=0$ calculations lead then to the same formation energies, 
except for the smallest simulation cell ($128$ lattice sites)  
in the case of the $\langle 111 \rangle$ cluster.
These formation energies have been also obtained with \abinitio calculations 
for a simulation cell containing $250$ lattice sites (Fig. \ref{fig:Ef_8sia111c15}).
Uncorrected $\bar{\varepsilon}=0$ calculations lead to an energy difference $\Delta E = -5.6$\,eV between the C15 and the $\langle 111 \rangle$ planar configuration, whereas this energy difference is only $\Delta E = -0.6$\,eV in $\sigma = 0$ calculations. This variation of the energy difference is rationalized once the elastic correction is added, 
and a good precision is obtained with this approach coupling \abinitio calculations and elasticity theory,
with an energy difference of $\Delta E = 3.5 \pm 0.2$\,eV.  
This elastic correction has been shown to accelerate the convergence of the point-defect formation and/or migration energies 
obtained from atomistic simulations, in particular from \abinitio calculations, 
in numerous other cases like SIA in hcp Zr \cite{Varvenne2013,Pasianot2016a}, 
vacancy in diamond silicon \cite{Varvenne2013}, or solute interstitials in bcc iron \cite{Souissi2016}.

\section{Conclusions}

Elasticity theory provides thus an efficient framework to model point-defects. 
Describing the point-defect as an equilibrated distribution of point-forces,
the long range elastic field of the defect and its interaction with other elastic fields 
are fully characterized by the first moment of this force distribution, 
a second rank symmetric tensor called the elastic dipole. 
This description is equivalent to an infinitesimal Eshelby inclusion
or an infinitesimal dislocation loop.
Knowing only the elastic constants of the matrix and the elastic dipole, 
a quantitative modeling of the point-defect and its interactions is thus obtained. 
The value of this elastic dipole can be either deduced from experimental data, 
like Vegard's law parameters, or extracted from atomistic simulations.
In this latter case, care must be taken to avoid finite-size effects, 
in particular for \abinitio calculations.
The definition through the residual stress appears as the most precise one to obtain the dipole tensors. 

The elastic description offers a convenient framework 
to bridge the scales between an atomic and a continuum description
so as to consider the interaction of the point-defects with various complex elastic fields.
This upscaling approach has already proven its efficiency in the modeling of elastodiffusion 
or in the calculation of absorption bias under irradiation. 
As the numerical evaluation of the elastic Green's function and its derivatives 
does not present nowadays any technical difficulty, 
such an elastic model offers also a nice route 
to simulate the evolution of a whole population of point-defects 
in a complex microstructure, considering their mutual interaction and their interaction
with other structural defects, in the same spirit as dislocation dynamics 
simulations are now routinely used to model the evolution of a dislocation microstructure.

\vspace{0.5cm}
\linespread{1}
\small

\textbf{Acknowledgements} -
This work was performed using HPC resources from GENCI-CINES and -TGCC (Grants 2017-096847).
The research was partly funded by the European Atomic Energy Community’s (Euratom) 
Seventh Framework Program FP7 under grant agreement No. 604862 (MatISSE project)  
and in the framework of the EERA (European Energy Research Alliance) Joint Program on Nuclear Materials.

\section*{References}
\bibliographystyle{elsarticle-num}
\bibliography{elasticity}

\end{document}